\documentclass{article}
\usepackage{arxiv}
\usepackage[utf8]{inputenc} % allow utf-8 input
\usepackage[T1]{fontenc}    % use 8-bit T1 fonts
\usepackage{hyperref}       % hyperlinks
\usepackage{url}            % simple URL typesetting
\usepackage{booktabs}       % professional-quality tables
\usepackage{amsfonts}       % blackboard math symbols
\usepackage{nicefrac}       % compact symbols for 1/2, etc.
\usepackage{microtype}      % microtypography
\usepackage{lipsum}		% Can be removed after putting your text content
\usepackage{natbib}
\usepackage{doi}
\usepackage{psfrag,color}
\usepackage[dvips]{graphicx}
\graphicspath{{./Figures/}} % declare the path(s) where your graphic files are
\DeclareGraphicsExtensions{.eps} % and their extensions to bypass within every instance of \includegraphics

\title{Sensor Management in Multi-Stage Stochastic Control Problems with Imperfect State Information}

\author{ {O.~Patrick Kreidl}\thanks{This manuscript met the term paper requirement for MIT Course~6.291 (Seminar on Systems, Communication, Control and Information) instructed by Professor Sanjoy Mitter in Fall~2002. Professor Mitter was beloved by students and colleagues the world over. He passed away June~26,~2023 at age~89; see {\tt https://news.mit.edu/2023/professor-emeritus-sanjoy-mitter-dies-0811}.} \\
	School of Engineering (with Courtesy Appointment in School of Computing)\\
	University of North Florida\\
	Jacksonville, FL 12224 \\
	\texttt{patrick.kreidl@unf.edu} \\
}

\date{}

\renewcommand{\headeright}{O.~Patrick Kreidl}
\renewcommand{\undertitle}{}
\renewcommand{\shorttitle}{Sensor Management in Multi-Stage Stochastic Control}

\nocite{*}

\begin{document}
\maketitle

\begin{abstract}
  Technological advancements in miniaturization and wireless
  communications are yielding more affordable and versatile sensors
  and, in turn, more applications in which a network of sensors can be
  actively managed to best support overall decision-making objectives.
  We propose modeling the opportunity for sensor management within
  multi-stage stochastic control problems with imperfect state
  information. Such formulations inherently assume the state of the
  modeled environment cannot be accessed directly but instead the
  controller can observe only noisy measurements of the state and,
  therefore, at each decision stage some form of state estimation is
  required before a control is actuated. The notion of sensor
  management arises when the modeled controls not only affect the
  subsequent evolution of the state but can also affect the nature of
  future measurements and, hence, the quality of state estimates that
  drive future control decisions. In principle, the optimal strategy
  for any appropriately modeled multi-stage stochastic control problem
  with imperfect state information (with or without opportunity for
  sensor management) is the solution to a dynamic program; in
  practice, the computational requirements are typically prohibitive
  yet dynamic programming methods are still useful to guide the
  development of effective suboptimal strategies.  In this spirit, we
  model the opportunity for sensor management within small-scale
  examples of two well-studied dynamic programming formulations,
  namely (1)~the finite-state/finite-action Partially-Observable
  Markov Decision Process (PO-MDP) and (2)~the
  Linear-Quadratic-Gaussian Regulator (LQGR). These examples admit
  solvable dynamic programs and confirm how the interplay between
  sensing and acting is a natural by-product of a dynamic programming
  solution.
\end{abstract}

\newpage
\section*{Ackowledgments}
\begin{quote}
``Better to remain silent and be thought a fool than to speak and remove all doubt'' \quad {\it ---Unknown, but I once heard Sanjoy Mitter say this circa 2005 upon being asked to comment during questions at the end of a MIT LIDS Colloquium, drawing chuckles from all in the room.}
\end{quote}

This manuscript was drafted in Fall 2002, the first semester of the author's return to MIT's Laboratory for Information and Decision Systems (LIDS) to begin doctoral studies as a member of Professor Alan Willsky's Stochastic Systems Group. It met the term paper requirement for MIT Course 6.291 (Seminar on Systems, Communication, Control and Information) instructed by Professor Sanjoy Mitter in Fall 2002. The self-selected topic, \emph{Sensor Management in Multi-Stage Stochastic Control Problems with Imperfect State Information}, marked the intellectual culmination of the author's coursework, teaching and research experiences since starting master's studies in Fall 1994.\footnote{The author received the S.M.~Degree in February 1996, with thesis titled {\it Distributed Cooperative Control Architectures for Automated Manufacturing Systems} co-advised by Professors Michael Athans and John Tsitsiklis, and eventually the Ph.D.~Degree in February 2008, with thesis titled {\it Graphical Models and Message-Passing Algorithms for Network-Constrained Decision Problems} advised by Professor Alan Willsky, both degrees in Electrical Engineering and Computer Science (EECS).} 

From Fall 1994 through Fall 1997, the author's time at MIT was supported by department-funded Teaching Assistantships (TAs) and grant-funded Research Assistships (RAs), spending Summer~1996 and Summer~1997 as an Intern at ALPHATECH, Inc (a research consulting firm in Burlington, MA). These coursework, teaching and research experiences fostered mathematical confidence, stimulated intellectual curiosity and, perhaps most importantly, provided access to outstanding mentorship:
\begin{description}
\item[Fall 1994:] Professor Munther Dahleh's instruction of 6.241 (Dynamic Systems), Professor Alan Willsky's instruction of 6.432 (Stochastic Processes, Detection and Estimation) and support via TA under Professor William Siebert's instruction of 6.003 (Signals and Systems).
\item[Spring 1995:] Professor Robert Gallager's instruction of 6.262 (Discrete Stochastic Processes), Professor Stanley Gershwin's instruction of 2.852 (Manufacturing Systems Analysis) and support via RA (funded by Schneider Industrial Controls, Inc.) under co-supervision of Professors Michael Athans, Dimitri Bertsekas and John Tsitsiklis.\footnote{Professor Michael Athans was incredibly encouraging of my ideas and efforts during this RAship, and in the years to follow gave me numerous additional opportunities to work and learn from the many interesting projects he was part of. As time went by, his personal advice became as important as his professional mentorship and I am forever grateful to have been part of his larger-than-life presence on Earth. He passed away May~26,~2020 at age~83; see {\tt https://news.mit.edu/2020/professor-emeritus-michael-athans-control-theory-pioneer-dies-0608}.}
\item[Fall 1995:] Professor Dimitri Bertsekas's instruction of 6.231 (Dynamic Programming and Stochastic Control) and support via TA under Professor Munther Dahleh's instruction of 6.003 (Signals and Systems), appointed Head TA due to prior year's TA experience with the course.
\item[Spring 1996:] Professor Richard Melrose's instruction of 18.100B (Analysis I) and support via TA under Professor Alvin Drake's instruction of 6.041 (Probabilistic Systems Analysis). Professors Alan Oppenheim and Alan Willsky included the author among the graduate students tasked to proof the chapters of their forthcoming textbook, the second edition of \emph{Signals and Systems}.
\item[Summer 1996:] Intern at ALPHATECH, Inc.,~co-supervised by Dr.~Leonard Lublin and Dr.~Mark Luettgen under the Modeling and Simulation Directorate, co-developing mathematical models, optimization methods and customized software for the prediction of, and course-of-action analysis in response to, catastrophic disruptions in international economic activity.
\item[Fall 1996:] Professor Thomas Magnanti's instruction of 6.251 (Intro to Mathematical Programming) and support via TA under Professor Alvin Drake's instruction of 6.041 (Probabilistic Systems Analysis), appointed Head TA due to prior semester's TA experience with the course.
\item[Spring 1997:] Professor Victor Guillemin's instruction of 18.103 (Fourier Analysis: Theory and Applications) and support via TA under Professor Munther Dahleh's instruction of 6.003 (Signals and Systems), again appointed Head TA. The MIT EECS Department selected me to receive the 1997 Carlton E.~Tucker Graduate Teaching Award. 
\item[Summer 1997:] Intern at ALPHATECH, Inc.,~co-supervised by Dr.~Joel Douglas under the Fusion Technologies Directorate, co-developing mathematical models, optimization methods and customized software for large-scale Bayesian Networks; also, by invitation of Dr.~Robert Tenney, Vice President of Technical Marketing, served as Student Member of the DARPA-sponsered Information Science and Technology (ISAT) Study Group on {\it The Dynamic Database}, a select pool of government officials, industry experts and university scholars charged to frame a twenty-year research and development program in support of next-generation military and civilian operations.
\end{description}

From 1997 to 2002, taking from MIT a Leave-of-Absence (in Good Standing), the author joined the newly-opened Arlington, Virgina office of ALPHATECH, Inc.~as a full-time Senior Analyst. From 1997 to 2000, under supervision of Dr.~Christoper J.~Donohue in the Optimization Directorate, he co-developed mathematical models, optimization methods and customized software for the automatic generation and synchronization of navigation plans and sensor schedules in multi-asset, air-to-ground surveillance problems. Professor Steven Marcus from the University of Maryland (UMD) in College Park served as an expert consultant, through which the team would grow to include UMD graduates Dr.~Craig Lawrence, Dr.~Andrew Newman and Dr.~Scott Laprise. From 2000 to 2002, under supervision of Dr.~Tiffany Frazier in the Advanced Computing Directorate, the author pioneered a control-theoretic framework for computer security problems and co-developed a host-based intrusion defense system to protect web-servers from automated, zero-day, Internet worms\footnote{This work received ALPHATECH's 2002 Joseph G.~Wohl Memorial Achievement Award and was subsequently both published (O.P. Kreidl and T.M. Frazier.``Feedback control applied to survivability: a host-based autonomic defense system'' \emph{IEEE Trans. on Reliability}, 53(1):148-166, Mar 2004.) and patented (T.M. Frazier and O.P. Kreidl.``Control systems and methods using a partially-observable Markov decision process'' U.S. Patent 7,363,515, Issued Apr 22, 2008).}. From 1998 to 2001, the author was also an Adjunct Professor of Electrical and Computer Engineering at George Mason University (his undergraduate Alma Mater) in Fairfax, VA, instructing ECE320 (Signals and Systems I) in Fall~1998, ECE421 (Classical Control Theory) in Spring~1999 and again ECE320 (Signals and Systems I) in Spring~2001. The Spring~2001 instruction included the administration of the \emph{Signals and Systems Concept Inventory (SCCI) Exam}, at that time under development via an NSF-sponsored grant led by Principal Investigators John R.~Buck (Dartmouth College), Kathleen E.~Wage (George Mason University), Cameron H.~G.~Wright (University of Wyoming) and Thad B.~Welch (U.S.~Naval Academy).\footnote{K.~E.~Wage, J.~R.~Buck,C.~H.~G.~Wright and T.~B.~Welch, ``The signals and systems concept inventory,'' \emph{IEEE Transactions on Education}, 48(3):448--461, August, 2005.}

Today, it is \today ~and the author keeps no ambition to conventionally publish this manuscript, so is choosing to upload it to the \emph{arXiv} open-access online repository. The manuscript is unreviewed, but for what it's worth based solely on the term paper submission Professor Mitter did in Fall 2002 assign an ``A'' for 6.291. Readers are invited to communicate questions or errata to the author via email: {\tt patrick.kreidl@unf.edu}

\newpage
\setlength{\topmargin}{-0.5in}
\setlength{\headsep}{0.5in}
\pagestyle{myheadings}
\addtocounter{page}{-1}
\markright{MIT-EECS 6.291 -- Fall 2002: Term Paper \hfill 
Kreidl--}

\tableofcontents

\newpage
\listoffigures
\listoftables

\newpage
\section{Introduction \label{Intro}}

\subsection{Motivation}
Any action-able choice to be made must strike a balance between the
desirability of timely making the most informed decision versus the
undesirability of expending resources (e.g., time, money) to gather
relevant information. For example, consider a physician who must
determine whether a patient's symptoms merit some subset of available
treatments.  If a treatment itself imposes some risk to the patient,
perhaps just in the long-term, the physician may find it prudent to
first perform certain diagnostic tests (e.g., blood work, X-ray),
delaying treatment until incorporating the results of the tests.  On
the other hand, certain symptoms may be of the nature where
inappropriate inaction carries serious short-term risks to the patient
and, thus, the physician chooses to immediately begin a treatment even
if some possibility exists that the treatment is unnecessary. In
deciding between treatments or tests or both, the physician
must carefully balance the immediate and future costs associated with
the selected response, factoring in the degrees of
uncertainty in both the current assessment of the patient's true
condition as well as the anticipated outcomes of the current decision.

The physician's situation is a typical instance of a general {\it
  multi-stage stochastic control problem with imperfect state
  information}.  In such problems, the essential information for the
purposes of decision-making, or the {\it state}, is not directly
accessible (e.g., the physician is ultimately interested in the
patient's true condition, or the root cause of the observable
symptoms).  Moreover, the actual outcome of a decision may not be
fully predictable yet can be anticipated to a certain extent; however,
it is certain that the ultimate objective (e.g., the physician cures
the patient) can only be achieved provided an action-able choice is
eventually made. The available controls at each decision stage can be
broadly classified into two categories: collect more state-related
information, hoping that the outcome will improve the
``state-of-knowledge'' for a future decision, or commit to a
particular action-able choice based on the current
``state-of-knowledge,'' hoping that the outcome will fall within
ultimate objectives. It is intuitively clear that a best decision
strategy should consider some coordinated mixture of gathering
information and committing to actions; of course, specifically
establishing the best such strategy, especially its dependence on the
various parameters of uncertainty, becomes non-trivial.

The situation is no less daunting on grander scales, but the
increasing reliance on information systems within critical military
and civilian operations coupled with the proliferation of data
collection devices, or {\it sensors}, motivates continued improvements
in automation that harnesses and processes collected data to support
more effective decision-making. Traditionally, most sensing
applications viewed the process of data collection as a strictly
passive activity, meaning a sensor's characteristics established
before its deployment persisted until the end of its lifetime. In the
future, due primarily to technological advancements in miniaturization
and wireless communications, more affordable and more versatile
sensors are being produced that are, in turn, yielding more
applications within which a network of sensors can be actively managed
to dynamically adapt the data collection process to ever-changing
environmental conditions and decision-making objectives.  The
potential benefits afforded by the opportunity for such {\it sensor
  management} assumes, as for the above small-scale example involving
the physician, that the decision strategy manages the flexible sensor
resources in a constructive manner with respect to achieving the
ultimate objectives.

\subsection{Outline and Scope}
This paper leverages discrete-time stochastic control theory as the
mathematical framework within which sensor management issues can be
modeled and analyzed. Section~\ref{DisStoCon} summarizes the key
methods and results of this generally applicable theory, introducing
both mathematical notation and key concepts that later sections of the
paper take for granted. The summary begins in mathematical generality
and presents successive simplifying concepts, identifying the
necessary assumptions assumed from step to step.  Sections~\ref{POMDP}
and \ref{LQGR} each present well-studied instances of a multi-stage
stochastic control problem with imperfect state information, the
Partially-Observable Markov Decision Process (PO-MDP) and the
Linear-Quadratic-Gaussian Regulator (LQGR), respectively, and in each
case uses a small-scale example to consider sensor management issues.
The analysis of Section~\ref{POMDP} is predominantly computational
whereas the analysis in Section~\ref{LQGR} is more analytical.
Section~\ref{Conclu} summarizes the points of this paper and their
relation to the management of modern sensor networks.

This paper does \underline{not} address several relevant components of
the overall sensor network management problem, namely the inherently
distributed architecture of the network and the induced communications
and decentralization issues. The modest focus of this paper is simply
to motivate a stochastic control formulation to crystallize sensor
management issues, demonstrating that the desired integration of the
sensing and acting resources is a natural by-product of a stochastic
control solution. Multi-stage stochastic control formulations already
lead to formidable optimization problems within the classical
centralized architectures---decentralization tends to make the
underlying optimization problems only more complex.

\section{Overview of Discrete-Time Stochastic Control \label{DisStoCon}}
General discrete-time control theory is a mature branch of mathematics
that addresses the modeling of multi-stage decision objectives within
a dynamic environment and the associated off-line design of a control
strategy that, once the controller is implemented on-line, promises to
best satisfy the stated objectives. Stochastic control theory further
assumes there is uncertainty in either the way the modeled state of
the environment evolves or the way it is observed or both, leading to
a reliance on probabilistic models and the off-line design of control
strategies that best satisfy the stated objectives in an on-average
sense. Because most realistic sensors measure quantities of interest
imperfectly, the desire to examine issues related to sensor management
enters the realm of a {\it discrete stochastic control problem with
  imperfect state information} \cite{Ber95:DPOC1}. The following subsections summarize
key concepts, methods and results for these kinds of problems in order
to both introduce notation and context that will prove useful in
later sections of this paper.

\subsection{Mathematical Models \label{MathModels}}
A general discrete-time stochastic control problem formulation involving
$K$ total decision stages relies on three key mathematical models
(expressed for each stage $k = 0, 1, \ldots, K-1$):
\begin{itemize}
\item a {\it system equation} $x_{k+1} = f_k(x_k,u_k,w_k)$, defining the
stochastic evolution of the {\it state} $x_k$ and its dependence on the
{\it control} $u_k$ and the random {\it disturbance} $w_k$ characterized
by a known conditional probability distribution $P_{w_k\mid x_k,
u_k}$;

\item a {\it cost equation} $g_k(x_k,u_k,w_k)$, quantifying the relative
  undesirability of all possible single-stage outcomes involving the
  current state $x_k$, selected control $u_k$ and random disturbance
  $w_k$; and

\item a {\it measurement equation} $z_k = h_k(x_k, u_{k-1}, v_k)$,
  defining the dependence of {\it measurement} $z_k$ on the current
  state $x_k$, the preceding control $u_{k-1}$ and the random {\it
  noise} $v_k$ characterized by a known conditional probability
  distribution $P_{v_k\mid x_k, {\cal I}_{k-1}}$ where we denote by 
  $$
  {\cal I}_{k-1} \subseteq
  \{x_{k-1},\ldots,x_0,u_{k-1},\ldots,u_0,w_{k-1},\ldots,w_0,v_{k-1},
  \ldots,v_0\}
  $$ 
  the set of variables from previous stages that
  statistically influence $v_k$.
\end{itemize}
A control $u_k$ is defined as any response available to the controller
with potential to causally influence the outcome of current or
future decision stages.  Note that, in the system equation, the state
$x_k$ is defined according to Markovian assumptions so that knowledge
of its current value implies any additional information about the past
is redundant or irrelevant for choosing future controls; moreover, the
distribution characterizing disturbance $w_k$ may only depend on the
current state and control.  These restrictions on the system equation
are not as strong as they initially appear, as problems that violate
such assumptions can typically be reformulated to the specified form
via {\it state augmentation}, or enlargement of the state space. Also
note that the measurements $z_k$ correspond to state-related
information that the controller can directly access---in the special
case of perfect state information, the measurement equation
degenerates to the trivial relation $z_k = x_k$.

Implicit in the above mathematical relations characterizing each stage
$k$ are specified sets of the possible states, controls, disturbances,
measurements and noises; denote by $S_k$ the {\it state space}, by
$C_k$ the {\it control space}, by $W_k$ the {\it disturbance space},
by $Z_k$ the {\it measurement space} and by $V_k$ the {\it noise
  space}. Then, more formally, the {\it system model} is the sequence
of mappings $f_k:S_k\times C_k\times W_k \rightarrow S_{k+1}$ and
distributions $P_{w_k\mid x_k,u_k}$; the {\it cost function} is
the sequence of mappings $g_k:S_k\times C_k\times W_k \rightarrow
\Re$; and the {\it measurement model} is the sequence of mappings
$h_k:S_k\times C_{k-1}\times V_k \rightarrow Z_k$ and distributions
$P_{v_k \mid x_k, {\cal I}_{k-1}}$. Also implicit are the
specialized forms of the models at the initial and terminal boundaries
of the multi-stage decision process. At initialization, where $k = 0$,
the initial state $x_0$ is characterized by a known probability
distribution $P_{x_0}$ over the space $S_0$ and the initial
measurement equation is more precisely stated as $z_0 = h_0(x_0,v_0)$
where $v_0$ is characterized by the known conditional distribution
$P_{v_0\mid x_0}$. At termination, where $k = K$, the
undesirability of the terminal state $x_K$ is quantified by a cost
$g_K(x_K)$.

Table~\ref{ModParams} lists all the components of a specified
discrete-time stochastic control model. The modeled process begins in
a random initial state $x_0$ realized according to distribution
$P_{x_0}$. Then, a random noise $v_0$ is realized according to
distribution $P_{v_0\mid x_0}$, generating the measurement $z_0$ via
equation $h_0$. The controller processes the measurement and selects a
control $u_0$ upon which a random disturbance $w_0$ is realized
according to distribution $P_{w_0\mid x_0,u_0}$, incurring a cost
$g_0$ and generating via equation $f_0$ the subsequent transition to
state $x_1$.  This state transition event simultaneously ends the 0th
decision stage and begins the next. The stages $k = 1, \ldots, K-1$
evolve in the same manner, only according to the stage-specific
equations and distributions, until the terminal state $x_K$ is
eventually reached and the associated terminal cost $g_K$ is incurred.

\begin{table}[ht]
\caption{Notation for a General Discrete-Time Stochastic Control Model}
\begin{center}
\begin{tabular}{ll} \hline \hline
SYMBOL & DESCRIPTION \\ \hline
$K$ & total number of decision stages \\[0.1in]
$S_k$ & state space in stage $k = 0, 1, \ldots, K$ \\
$C_k$ & control space in stage $k = 0, 1, \ldots, K-1$ \\
$W_k$ & disturbance space in stage $k = 0, 1, \ldots, K-1$ \\
$Z_k$ & measurement space in stage $k = 0, 1, \ldots, K-1$ \\
$V_k$ & noise space in stage $k = 0, 1, \ldots, K-1$ \\[0.1in]
$x_k \in S_k$ & state in stage $k$ (initial state $x_0$ random with distribution $P_{x_0})$\\
$u_k \in C_k$ & control in stage $k$ \\
$w_k \in W_k$ & disturbance in stage $k$ (random with distribution 
                $P_{w_k\mid x_k u_k}$) \\
$z_k \in Z_k$ & measurement in stage $k$ \\
$v_k \in V_k$ & noise in stage $k$ (random with distribution 
                $P_{v_k\mid x_k,{\cal I}_{k-1}}$ where \\
              & $\qquad {\cal I}_{k-1} \subseteq
                \{x_{k-1},\ldots,x_0,u_{k-1},\ldots,u_0,w_{k-1},
                \ldots,w_0,v_{k-1},\ldots,v_0\}$) \\[0.1in]
$x_{k+1} = f_k(x_k,u_k,w_k)$ & system equation at stage $k=0,1,\ldots,K-1$ 
           ($f_k:S_k\times C_k \times W_k \rightarrow S_{k+1}$) \\
$g_k(x_k,u_k,w_k)$ & cost equation at stage $k=0,1,\ldots,K-1$ 
           ($g_k:S_k\times C_k \times W_k \rightarrow \Re$) \\
$g_K(x_K)$ & cost equation at terminal stage $K$ \\
$z_k = h_k(x_k,u_{k-1},v_k)$ & measurement equation at stage $k=0,1,\ldots,K-1$ 
($h_k:S_k\times C_{k-1} \times V_k \rightarrow Z_k$)  \\ \hline \hline
\end{tabular}
\end{center}
\label{ModParams}
\end{table}

Given the modeled decision stages, questions of controllability,
observability and stability are fundamental system-level concerns prior
to embarking on the design of a decision
strategy. Qualitatively\footnote{We forego the mathematical definitions
of these fundamental concepts because they are not elaborated upon nor
applied any further in subsequent sections.}, {\it controllability}
refers to a measure of how effective the controls are with respect to
altering the true state. If the system is uncontrollable, applying
control has no predictable influence on how the future evolves from its
current state. Complete controllability, on the other hand, implies it
is possible through some finite sequence of controls to alter the
current state to any other desired state. Similarly, {\it observability}
refers to a measure of how informative the sensor measurements are with
respect to estimating the true state. If the system is unobservable, the
sensor measurements have no useful information content to help infer the
true state. Complete observability, on the other hand, implies it is
possible through some finite sequence of measurements to uniquely
determine the true state. Finally, {\it stability} refers to a system
which, in the absence of external input, eventually settles to some
equilibrium state. When stability of the controlled system is not
satisfied, the immediate objective must be to stabilize the system and
only upon stabilization is it possible to proceed with objectives to
improve system performance. Provided stability guarantees are preserved,
the cost function allows the performance problem to be stated in a
manner well-suited for mathematical optimization tools, where the
optimal performance is limited only by the inherent observability and
controllability properties of the underlying models\footnote{This
statement implicitly assumes the system/measurement models are
sufficiently accurate descriptions of the environment. Designing
controllers that are robust to modeling errors is an
important problem we deem beyond this paper's scope.}.

\subsection{Dynamic Programming}
In a dynamic and uncertain multi-stage process, a single-stage decision
cannot be viewed in isolation. Rather, the desire for low immediate cost
must be balanced against the undesirability of high expected future
costs. Numerical optimization via {\it dynamic programming}
mathematically captures this trade-off, explicitly accounting for all
modeled uncertainty in the process. The general technique relies on the
models defined in the previous subsection (see Table~\ref{ModParams})
and, in principle, provides an algorithm that terminates with a
minimum-cost, or {\it optimal}, solution. A dynamic program distinctly
recognizes the sequential nature of the problem so that the solution
generated off-line will exploit {\it information feedback}, allowing
each decision during on-line operation to depend on the most recent
information available from all observable data.  While the off-line
algorithm is nearly always computationally prohibitive, especially under
the assumption of imperfect state information, the mathematical
framework retains a practical value via the insight it provides to
determine an appropriate balance between problem representation and
solution computation and, ultimately, guide the development of tractable
yet effective {\it sub-optimal} strategies.

\subsubsection{Strategy Optimization}
Given $K$ modeled decision stages as listed in Table~\ref{ModParams},
consider first the objective of selecting a {\it control trajectory}
${\bf u}_K = (u_0, u_1, \ldots, u_{K-1})$ so as to minimize an {\it
  additive}\footnote{Similar to the remarks made about the specified
  form of the system equation in the preceding subsection, the specified
  additive assumption on the total cost is also not as restrictive as
  it appears---again via state augmentation, a non-additive total cost
  can generally be reformulated to the form in (\ref{eq:OLCost}).} 
{\it expected  total cost}
\begin{eqnarray}
G\left({\bf u}_K \mid P_{x_0}, \{ f_k, P_{w_k \mid
x_k, u_k}; k = 0, 1, \ldots, K-1\}
\right) = E\left[g_K(x_K) + \sum_{k=0}^{K-1} g_k(x_k,u_k,w_k)\right]
\label{eq:OLCost}
\end{eqnarray}
where the notation emphasizes that the expectation $E[\bullet]$ is only
well-defined once all aspects of the system model are specified. Note
that, for any specified control trajectory ${\bf u}_K$, it is only upon
combining the successive functions $f_0, \ldots, f_{K-1}$ with
probability distributions $P_{x_0}$ and $P_{w_0\mid x_0,
u_0}, \ldots , P_{w_{K-1}\mid x_{K-1}, u_{K-1}}$ that the
quantities $w_0, x_1, w_1, x_2 \ldots, w_{k-1}, x_K$ become a
succession of well-defined random variables. Implied by the system model
is a coupling between present and future decision stages, in the sense
that each control $u_k$ must be selected in a manner that anticipates
both an immediate and future influence on the state evolution, and
(\ref{eq:OLCost}) quantifies a balance between the associated immediate
and future expected costs.

Equation \ref{eq:OLCost} neglects to articulate an important aspect of a
sequential decision process, referred to as {\it information feedback},
which assumes that each control $u_k$ can be selected with some
knowledge (and ideally perfect knowledge) of the current state
$x_k$. Conceptually, {\it imperfect state information} problems are
treated no differently from {\it perfect state information}
problems---instead of relying on the true state $x_k$ that is assumed to
summarize all {\it essential} information for selecting the $k$th
control, there is a ``state-of-information'' derived from all received
measurements $z_0, z_1, \ldots, z_k$ and past controls $u_0, u_1,
\ldots, u_{k-1}$ that summarizes all the {\it observable}
information for selecting the $k$th control. Mathematically,
(\ref{eq:OLCost}) is generalized to depend on a {\it control strategy}
$\bar{\pi}_K$, or a trajectory of functions $(\bar{\mu}_0, \bar{\mu}_1,
\ldots, \bar{\mu}_{K-1})$ where, at each stage $k$, the available {\it
information vector} $I_k = (z_0, u_0, z_1, \ldots, u_{k-1}, z_k) =
(I_{k-1}, u_{k-1}, z_k )$ feeds the $k$th {\it control policy}
$\bar{\mu}_k$ that maps to the control according to $u_k =
\bar{\mu}_k(I_k)$. Thus, the optimization objective is more precisely
expressed as selecting control strategy $\bar{\pi}_K$ so as to minimize
\begin{eqnarray}
\begin{array}{ll}
\bar{J}\left(\bar{\pi}_K \mid P_{x_0}, \{h_k, P_{v_k\mid x_k,
  {\cal I}_{k-1}},f_k,P_{w_k \mid x_k, u_k}; k = 0, 1, \ldots,
  K-1\} \right) & = \qquad \qquad \qquad \qquad \\[0.1in]
\multicolumn{2}{r}{\displaystyle E\left[g_K(x_K) + \sum_{k=0}^{K-1}
  g_k(x_k,\bar{\mu}_k(I_k),w_k)\right]}
\end{array}
\label{eq:CLCost}
\end{eqnarray}
where the notation emphasizes that the expectation $E[\bullet]$ is now
only well-defined once all aspects of the system model and the
measurement model are specified.

The optimization expressed by (\ref{eq:CLCost}) versus that expressed
by (\ref{eq:OLCost}) contrasts the notions of, respectively, {\it
  closed-loop} versus {\it open-loop} control selection. The minimum
total expected cost achievable with a closed-loop formulation will
never exceed that achievable with the open-loop formulation and,
typically, the closed-loop optimization yields a significant reduction
in the expected total cost. This is because the off-line development
of a closed-loop strategy specifically accounts for the extra
state-related information that will become available as the on-line
decision process unfolds, allowing the closed-loop controller to adapt
appropriately to less expected events. Unfortunately, the on-line
benefits of a closed-loop controller come at the price of greatly
increased off-line complexity---instead of optimization over the space
of a variable sequence ${\bf u}_K$, the optimization occurs over the
far larger space of a function sequence $\bar{\pi}_K$. Nonetheless,
dynamic programming is the only general approach for sequential
optimization under uncertainty when striving to develop a closed-loop
strategy. It is in fact only via a dynamic programming approach that
the inherent difficulty of sequential decision-making under
uncertainty is rigorously exposed.

The main dynamic programming algorithm constructs an optimal policy in a
piecemeal fashion, beginning with the ``tail subproblem'' (i.e.,
considering just the last decision stage and its impact on the terminal
cost) and then recursively working backward.  In this way, each stage
of the cost minimization expressed over the space of policies
(functions) can get decomposed into information-dependent minimizations
over the space of controls (variables): \\
\begin{minipage}[h]{\textwidth}
\rule{\textwidth}{0.01in}
\begin{footnotesize}
{\bf Main Algorithm:} For every initial information vector $I_0$, the
  optimal cost is equal to $\bar{J}_0(I_0)$, where the function
  $\bar{J}_0$ is given by the last step of the following algorithm:

\begin{description}
\item[\it Initialization:] Given $I_{K-1}$ and a prospective control
  $u_{K-1}$, the following expectation $E[\bullet]$ occurs with respect to the
  joint distribution $P_{x_{K-1},w_{K-1}\mid I_{K-1}, u_{K-1}}$
  implied by the system model and measurement model:
\begin{eqnarray*}
\bar{J}_{K-1}(I_{K-1}) & = & \min_{u_{K-1} \in C_{K-1}} \left\{ E\left[
  g_{K-1}(x_{K-1}, u_{K-1}, w_{K-1}) + \right. \right. \\ & & \qquad
  \qquad \qquad \qquad \left. \left. g_K(f_{K-1}(x_{K-1}, u_{K-1},
  w_{K-1})) \mid I_{K-1}, u_{K-1} \right] \right\}
\end{eqnarray*}

\item[\it Recursion:] For stages $k = K-2, K-3, \ldots, 0$, given
  $I_k$ and a prospective control $u_k$, the following expectation
  $E[\bullet]$ occurs with respect to the joint distribution
  $P_{x_k,w_k,z_{k+1}\mid I_k, u_k}$ implied by the system model and
  measurement model:
\begin{eqnarray*}
\bar{J}_k(I_k) = \min_{u_k \in C_k} \left\{ E\left[ g_k(x_k, u_k, w_k) +
  \bar{J}_{k+1}(I_k,u_k,z_{k+1}) \mid I_k, u_k \right] \right\}
\end{eqnarray*}
\end{description}
At each iteration $k$, the minimization that yields $\bar{J}_k$ for each
$I_k$ simultaneously defines the associated optimal policy
$\bar{\mu}^*_k$; that is, if $u_k^* = \bar{\mu}^*_k(I_k)$ minimizes
$\bar{J}_k(I_k)$ for each possible $I_k$, then the resulting optimal
strategy is $\bar{\pi}^*_K = (\bar{\mu}^*_0,\bar{\mu}^*_1, \ldots,
\bar{\mu}^*_{K-1})$. \\[-0.1in]
\end{footnotesize}
\rule{\textwidth}{0.01in}
\end{minipage}
Note that each function $\bar{J}_k$ is usually called the {\it optimal
cost-to-go function} at time $k$ because, for any possible realization
of information vector $I_k$, it represents the total expected cost for
the remaining $(K-k)$-stage problem under optimal
decision-making. Relating back to (\ref{eq:CLCost}), the optimal cost
associated with closed-loop strategy $\bar{\pi}^*_K$ is then
\begin{eqnarray*}
\bar{J}\left(\bar{\pi}^*_K \mid P_{x_0}, \{h_k, P_{v_k\mid x_k,
  {\cal I}_{k-1}},f_k,P_{w_k \mid x_k, u_k}; k = 0, 1, \ldots,
  K-1\} \right) = E\left[\bar{J}_0(z_0)\right]
\end{eqnarray*}
where we recall that $I_0 = z_0$ and note that the initial state
distribution and stage $0$ of the measurement model imply a
well-defined distribution $P_{x_0,z_0}$ to compute the expectation
$E[\bullet]$.

\begin{figure}[ht]
\begin{center}
\psfrag{f_k(x_k,u_k,w_k)}[cc][cc]{$f_k(x_k,u_k,w_k)$}
\psfrag{h_k(x_k,u_k,v_k)}[cc][cc]{$h_k(x_k,u_{k-1},v_k)$}
\psfrag{(I_k,u_k,z_k)}[cc][cc]{$(I_{k-1},u_{k-1},z_k)$}
\psfrag{mu_k(I_k)}[cc][cc]{$\bar{\mu}_k(I_k)$}
\psfrag{x_k}[cc][cc]{$x_k$}
\psfrag{w_k}[cc][cc]{$w_k$}
\psfrag{v_k}[cc][cc]{$v_k$}
\psfrag{u_k}[cc][cc]{$u_k$}
\psfrag{z_k}[cc][cc]{$z_k$}
\psfrag{u_k-1}[cc][cc]{$u_{k-1}$}
\psfrag{I_k}[cc][cc]{$I_k$}
\psfrag{I_k-1}[cc][cc]{$I_{k-1}$}
\includegraphics{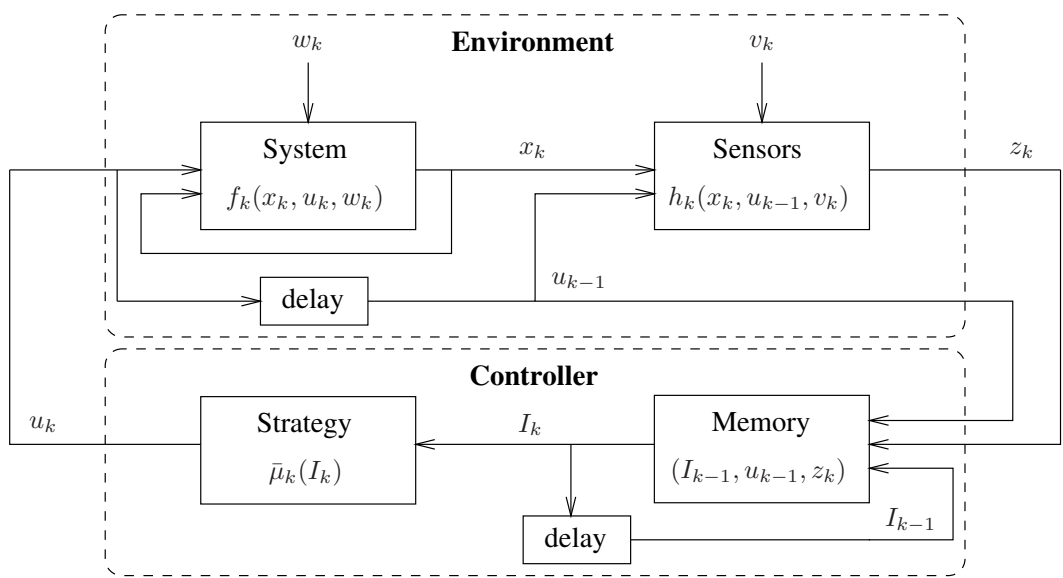}
\end{center}
\caption{On-line Architecture and Data Flow for a Closed-Loop Control System}
\label{fig:genArch}
\end{figure}

Figure~\ref{fig:genArch} summarizes the functional architecture and
on-line data flows implied by the mathematical models and the dynamic
programming solution described above. The system model $f_k$ and
measurement model $h_k$ together characterize each stage of the
stochastic environment, where at each stage of on-line operation the
measurement $z_k$ and preceding control $u_{k-1}$ are the only data
observable by the controller. This data gets stored into the
controller's memory and the entire observed history $I_k$ drives the
closed-loop control strategy $\bar{\pi}_K$, defined at each stage by
feedback policy $\bar{\mu_k}(I_k)$.  Note that the cost function
is not explicitly relevant in the on-line operation; of course, it
influences the off-line optimization so as to yield desirable control
strategies for the specified modeled environment.

\subsubsection{Separation Principle}
The {\it separation principle} is one of the most celebrated concepts
for stochastic control problems with imperfect state information.  It
establishes technical conditions whereby the on-line control strategy
can be decomposed into a data-driven {\it estimator} followed by an
estimate-driven {\it actuator} without loss of performance.  It is
perhaps surprising that the general models described in Subsection
\ref{MathModels} do in fact meet the conditions for controller
separation (see \cite{Wit71:SepEC}). It turns out that the main factor in
whether the separation principle applies is the {\it information
  pattern}, or the data that is assumed available as arguments to the
controller at the time a decision is to be made. In the context
of~\cite{Wit71:SepEC}, we have assumed a {\it strictly classical
  pattern}, which involves only a single {\it observation post}
co-located at a single {\it control station} (i.e., a centralized
architecture). The separation principle then applies where the
estimate at each stage is the {\it probabilistic state} $P_{x_k\mid
  I_k}$, or the conditional distribution of the state based on all
observed information.
 
Intuitively, the probabilistic state estimate can be viewed as the
``state-of-information,'' the result of optimally processing all
observed raw data into a mathematical form that is common to every
stage of the control strategy. To develop some insight into how this
separation arises, consider a single recursion of the main algorithm
\begin{eqnarray}
\bar{J}_k(I_k) = \min_{u_k \in C_k} \left\{ E\left[ g_k(x_k, u_k, w_k) +
  \bar{J}_{k+1}(I_k,u_k,z_{k+1}) \mid I_k, u_k \right] \right\}
\label{eq:DPeqn}
\end{eqnarray}
and recall that the expectation $E[\bullet]$ is computed over the
joint distribution $P_{x_k,w_k,z_{k+1}\mid I_k, u_k}$ that can be
expressed as the product $P_{w_k,z_{k+1}\mid I_k,x_k,u_k} \cdot P_{x_k
  \mid I_k,u_k}$. Let us take the off-line design of the optimal
cost-to-go function $\bar{J}_k$ for granted. Then, exploiting the Law
of Iterated Expectations (i.e., $E[A] = E\left[E[B|A]\right]$ for
random variables $A$ and $B$), the corresponding optimal policy at
each stage $k$ can be realized via an on-line computation
\[
\bar{\mu}_k^*(I_k) = \arg \min_{u_k \in C_k} \left\{ E\left[ E\left[
      g_k(x_k,u_k, w_k) + \bar{J}_{k+1}(I_k,u_k,z_{k+1}) \mid x_k,
      I_k, u_k \right] \mid I_k,u_k \right] \right\} \quad ,
\]
where the inner expectation occurs over the distribution
$P_{w_k,z_{k+1}\mid I_k,x_k,u_k}$ and the outer expectation occurs
over $P_{x_k\mid I_k,u_k}$. However, upon fixing a control strategy,
each control variable $u_k = \bar{\mu}_k(I_k)$ is completely
determined by the realized information vector $I_k$ and, therefore,
the random variable $u_k$ offers no additional information about
random variable $x_k$, implying $P_{x_k\mid I_k,u_k} = P_(x_k\mid I_k
\bar{\mu}_k(I_k)) = P_{x_k\mid I_k}$. Thus, an equivalent on-line
realization of the optimal control strategy is
\begin{eqnarray}
\bar{\mu}_k^*(I_k) & = & \arg \min_{u_k \in C_k} 
\left\{ E\left[ E\left[ g_k(x_k,u_k, w_k) + 
\bar{J}_{k+1}(I_k,u_k,z_{k+1}) \mid x_k, I_k, u_k \right] \mid
  I_k \right] \right\} \quad , 
\label{eq:DPeqnsep}
\end{eqnarray}
where the inner expectation still occurs over the distribution
$P_{w_k,z_{k+1}\mid I_k,x_k,u_k}$ but the outer expectation now occurs
over $P_{x_k\mid I_k}$.

\begin{figure}[ht]
\begin{center}
\psfrag{(I_k,u_k,z_k)}[cc][cc]{$(I_{k-1},u_{k-1},z_k)$}
\psfrag{mu_k(I_k)}[cc][cc]{$\mu_k(P_{x_k\mid I_k})$}
\psfrag{nu_k(I_k)}[cc][cc]{$\bar{\nu}_k(I_k)$}
\psfrag{P_k}[cc][cc]{$P_{x_k\mid I_k}$}
\psfrag{u_k}[cc][cc]{$u_k$}
\psfrag{z_k}[cc][cc]{$z_k$}
\psfrag{u_k-1}[cc][cc]{$u_{k-1}$}
\psfrag{I_k}[cc][cc]{$I_k$}
\psfrag{I_k-1}[cc][cc]{$I_{k-1}$}
\includegraphics{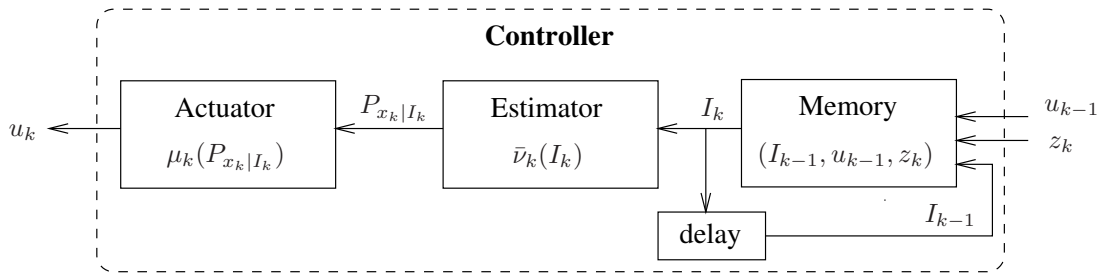}
\end{center}
\caption{On-Line Controller Architecture Under the Separation Principle} 
\label{fig:genArchSep}
\end{figure}

Provided that the distribution $P_{x_k\mid I_k}$ has been computed,
the outer expectation in (\ref{eq:DPeqnsep}) being over a distribution
that is independent of control $u_k$ implies that, for any policy of
the form $\bar{\mu}_k(I_k)$, there is also a function
$\mu_k(P_{x_k\mid I_k})$ that maps to the equivalent control and
achieves the equivalent cost-to-go. In other words, as illustrated in
Fig.~\ref{fig:genArchSep}, the on-line implementation of each control
policy $u_k = \bar{\mu}_k(I_k)$ can be decomposed into two successive
function evaluations: an {\it estimation policy} of the form
$P_{x_k\mid I_k} = \bar{\nu}_k(I_k)$, mapping the information vector
to the probabilistic state estimate, followed by an {\it actuation
  policy} of the form $u_k = \mu_k(P_{x_k\mid I_k})$, mapping the
estimate to the desired control. Thus, the closed-loop control
strategy $\bar{\pi}_K$ can be conveniently viewed as the composite of
an {\it estimation strategy} $\bar{\phi}_K = (\bar{\nu}_0,
\bar{\nu}_1, \ldots, \bar{\nu}_{K-1})$ and an {\it actuation strategy}
$\pi_K = (\mu_0, \mu_1, \ldots, \mu_{K-1})$, in terms of which the
closed-loop optimization objective in (\ref{eq:CLCost}) becomes
\begin{eqnarray}
\begin{array}{ll}
J\left(\bar{\phi}_K, \pi_K \mid P_{x_0}, \{h_k,
  P_{v_k\mid x_k, {\cal I}_{k-1}},f_k,P_{w_k \mid x_k, u_k}; k =
  0, 1, \ldots, K-1\} \right) & = \qquad \qquad \qquad \\[0.1in]
  \multicolumn{2}{r}{\displaystyle E\left[g_K(x_K) + \sum_{k=0}^{K-1}
  g_k(x_k,\mu_k(\bar{\nu}_k(I_k)),w_k)\right]}
\end{array}
\label{eq:CLCostSep}
\end{eqnarray}

The off-line determination of each estimation policy $\bar{\nu}_k(I_k)$
is called the {\it filtering problem} and, while a difficult problem in
general, the separation principle suggests that its solution depends on
the system model up through stage $k-1$ and the measurement model up
through stage $k$ but it is entirely independent of subsequent stages of
the models as well as the overall cost function and, perhaps most
surprising, the overall actuation strategy! Concerning the optimal
actuation strategy $\pi_K$, it firstly takes for granted the solution to
the filtering problem and, secondly, (\ref{eq:DPeqnsep})
implies that it's off-line solution is no less computationally difficult
than the off-line solution for the optimal control strategy without
separation, $\bar{\pi}_K$. In particular, as long as the distribution
over which the inner expectation is computed retains a dependence on the
vector $I_k$, even upon conditioning on the current state $x_k$, the
recursive calculation of the cost-to-go function $\bar{J}_k$ also
retains the explicit dependence on $I_k$. The following subsection
discusses how certain model restrictions, in particular assumptions on
the extent to which the noise distribution $ P_{v_k\mid x_k,{\cal
I}_{k-1}}$ depends on variables from past stages (i.e., restrictions on
the generality of the set ${\cal I}_{k-1}$), admit the separation
principle to be of far greater practical significance with respect to
the off-line determination of the estimation and actuation strategies
and, thus, the optimal closed-loop controller.

\subsubsection{Recursive Estimation}
The primary difficulty with executing the algorithm stated in
section~\ref{DisStoCon} is that it involves the full information
vector $I_k$, a space of expanding dimension as $k$ increases.
Moreover, as conveyed by Figs.~\ref{fig:genArch} and
\ref{fig:genArchSep}, the associated implication for the on-line
controller is that unlimited memory is required as $k$ increases.  The
discussion surrounding Fig.~\ref{fig:genArchSep} recognized an on-line
separation of the optimal control strategy into an information-driven
estimator followed by an estimate-driven actuator; however, as
fore-shadowed in the preceding paragraph, an analogous simplification
of the off-line strategy optimization does not materialize without
additional restrictions on the measurement model.

By inspection of (\ref{eq:DPeqnsep}), as long as the distribution
$P_{w_k,z_{k+1}\mid I_k,x_k,u_k}$ over which the inner expectation is
computed retains a dependence on the vector $I_k$, even upon
conditioning on the current state $x_k$, the recursive calculation of
the cost-to-go function $\bar{J}_k$ also retains the explicit
dependence on $I_k$. This joint distribution can be expressed as the
product $P_{z_{k+1}\mid I_k,x_k,u_k,w_k} \cdot P_{w_k\mid
  I_k,x_k,u_k}$ and, furthermore, by definition of the system model we
know $w_k$ can depend only on $x_k$ and $u_k$ implying $P_{w_k\mid
  I_k,x_k,u_k} = P_{w_k\mid x_k,u_k}$. Thus, it follows that the
explicit dependence on $I_k$ arises in the main dynamic programming
algorithm strictly due to its influence on the distribution
characterizing future measurement $z_{k+1}$---in other words, we wish
to establish conditions such that $P_{z_{k+1}\mid I_k,x_k,u_k,w_k} =
P_{z_{k+1}\mid x_k,u_k,w_k}$. Noticing that the distribution
$P_{z_{k+1}\mid I_k,x_k,u_k,w_k}$ is derived from the measurement
equation $z_{k+1} = h_{k+1}(x_{k+1},u_k,v_{k+1})$ and noise
distribution $P_{v_{k+1}\mid x_{k+1},{\cal I}_k}$, the desired
conditions are for the noise $v_{k+1}$ to depend at most on the
immediately preceding state, control and disturbance $x_k, u_k, w_k$.

In short, we have just argued that if the measurement model specified in
Table~\ref{ModParams} restricts the set ${\cal I}_{k-1} \subseteq
\{x_{k-1},u_{k-1},w_{k-1}\}$ for every $k$, then (\ref{eq:DPeqnsep})
simplifies to
\begin{eqnarray}
\bar{J}_k(I_k) & = & \min_{u_k \in C_k} \left\{ E\left[ E\left[ g_k(x_k,
  u_k, w_k) + \bar{J}_{k+1}(I_k,u_k,z_{k+1}) \mid x_k, u_k \right] \mid
  I_k \right] \right\}
\label{eq:DPeqnrec}
\end{eqnarray}
where the outer expectation still occurs over the probabilistic state
$P_{x_k\mid I_k}$ but now the inner expectation occurs over
$P_{w_k,z_{k+1}\mid x_k,u_k}$. Given the system model at stage $k$ and
the measurement model at stage $k+1$ and taking the filtering problem
for granted, (\ref{eq:DPeqnrec}) simplifies to
\begin{eqnarray*}
\bar{J}_k(I_k) = \min_{u_k \in C_k} Q_k(P_{x_k\mid I_k}, u_k) 
\equiv J_k\left(P_{x_k\mid I_k}\right)
\end{eqnarray*}
for a suitable function $Q_k$ and a cost-to-go function $J_k$ that
takes the probabilistic state as its argument rather than the entire
information vector. Thus, by restricting the sensor model, each
probabilistic state estimate $P_{x_k\mid I_k}$, essentially a
processed form of the information vector $I_k$, is equivalent in
information content as far as selecting control $u_k$ is concerned, a
statistical property defined as {\it sufficient}. Armed with the
probabilistic state $P_{x_k\mid I_k}$ as a sufficient statistic, the
main algorithm can be reformulated so that the estimation strategy is
a recursive operation defined by the system and measurement models and
the actuation strategy takes the probabilistic
state as its argument: \\
\begin{minipage}[h]{\textwidth}
\rule{\textwidth}{0.01in}
\begin{footnotesize}
{\bf Main Algorithm With Recursive Estimation:} Let the noise
distribution of the measurement model be restricted such that the set
${\cal I}_{k-1} \subseteq \{x_{k-1},u_{k-1},w_{k-1}\}$.
\begin{description}
\item[\sc Estimation Strategy $\phi_K$:] Sufficiency implies that the
  filtering problem admits a recursive solution, meaning we can first
  compute $P_{x_0\mid I_0} = \nu_0(I_0) = P_{x_0\mid z_0}$ and then
  for $k > 0$ compute
\begin{eqnarray*}
P_{x_k\mid I_k} = \bar{\nu}_k(I_k) = \nu_k(P_{x_{k-1}\mid I_{k-1}},
u_{k-1}, z_k)
\end{eqnarray*}
where each $\nu_k$ is some function that, in principle, can be
determined directly from the system model at stage $k-1$ and the
measurement model at stage $k$.  The sequence of functions $\phi_K =
\{ \nu_0, \nu_1, \ldots, \nu_{K-1}\}$ defines the (recursive) estimation
strategy.
  
\item[\sc Actuation Strategy $\pi^*_K$:] For every initial probabilistic
  state $P_{x_0\mid I_0}$, the optimal cost equals $J_0(P_{x_0\mid
  I_0})$, where the function $J_0$ is given by the last step of the
  following algorithm:

\begin{description}
\item[\it Initialization:] At stage $K-1$, define
\begin{eqnarray*}
Q_{K-1}(P_{x_{K-1}\mid I_{K-1}},u_{K-1}) & = & 
E\left[ E\left[ g_{K-1}(x_{K-1}, u_{K-1}, w_{K-1}) + \right. \right. \\
& & \qquad \qquad \left. \left. g_K(f_{K-1}(x_{K-1}, u_{K-1}, w_{K-1}) 
\mid x_{K-1}, u_{K-1} \right] \mid I_{K-1} \right]
\end{eqnarray*}
where the inner expectation $E[\bullet]$ is taken with respect to the
specified disturbance distribution $P_{w_{K-1}\mid x_{K-1},u_{K-1}}$
and the outer expectation is taken with respect to the probabilistic
state $P_{x_{K-1}\mid I_{K-1}}$. Then, the cost-to-go can be expressed
as
\begin{eqnarray*}
J_{K-1}(P_{x_{K-1}\mid I_{K-1}}) = \min_{u_{K-1} \in C_{K-1}}
  Q_{K-1}(P_{x_{K-1}\mid I_{K-1}},u_{K-1}) \quad .
\end{eqnarray*}
  
\item[\it Recursion:] For stages $k = K-2, K-3, \ldots, 0$, define
\begin{eqnarray*}
Q_k(P_{x_k\mid I_k},u_k) = E\left[ E\left[ g_k(x_k, u_k, w_k) + 
J_{k+1}(\nu_{k+1}(P_{x_k\mid I_k}, u_k,z_{k+1}))\mid x_k, u_k \right] 
\mid I_k \right]
\end{eqnarray*}
where the inner expectation $E[\bullet]$ occurs over the joint
distribution $P_{w_k,z_{k+1}\mid x_k,u_k}$ implied by the system model
at stage $k$ and the measurement model at stage $k+1$ and the outer
expectation occurs over the probabilistic state $P_{x_k\mid I_k}$.
Then,
\begin{eqnarray*}
J_k(P_{x_k\mid I_k}) = \min_{u_k \in C_k} Q_k(P_{x_k\mid I_k},u_k) \qquad .
\end{eqnarray*}
\end{description}
At each iteration $k$, the minimization that yields $J_k$ for each
${\cal P}_{x_k\mid I_k}$ simultaneously defines the associated optimal
policy $\mu^*_k$; that is, in terms of the functions $Q_k$, the
optimal stage-$k$ actuation policy is given by
\begin{eqnarray*}
\mu_k^*(P_{x_k\mid I_k}) = \arg \min_{u_k \in C_k} 
Q_k(P_{x_k\mid I_k},u_k)
\end{eqnarray*}
and the resulting actuation strategy $\pi^*_K = (\mu^*_0,\mu^*_1,
\ldots, \mu^*_{K-1})$ is optimal. 
\end{description}
The optimal closed-loop control strategy $\bar{\pi}^*_K$ is the
composite of the estimation strategy $\phi_K$ and optimal actuation
strategy $\pi_K^*$. \\[-0.1in]
\end{footnotesize}
\rule{\textwidth}{0.01in}
\end{minipage}
Relating back to (\ref{eq:CLCostSep}), the optimal cost associated
with the composite closed-loop strategy $\bar{\pi}^*_K$ is then
\begin{eqnarray}
J\left(\phi_K, \pi^*_K \mid P_{x_0}, \{h_k,
  P_{v_k\mid x_k, {\cal I}_{k-1}},f_k,P_{w_k \mid x_k, u_k}; k =
  0, 1, \ldots, K-1\} \right) = E\left[J_0(P_{x_0\mid z_0})\right]
\label{eq:CLCostRec}
\end{eqnarray}
where, as before, we recall that $I_0 = z_0$ and note that the
measurement model and initial state distribution imply a well-defined
distribution ${\cal P}_{x_0,z_0}$ to compute the expectation
$E[\bullet]$.

\begin{figure}[ht]
\begin{center}
  \psfrag{(I_k,u_k,z_k)}[cc][cc]{$(I_{k-1},u_{k-1},z_k)$}
  \psfrag{mu_k(I_k)}[cc][cc]{$\mu_k(P_{x_k\mid I_k})$}
  \psfrag{nu_k(P_k-1,u_k-1,z_k)}[cc][cc]{$\nu_k(P_{x_{k-1}\mid
      I_{k-1}}, u_{k-1}, z_k)$} \psfrag{P_k}[cc][cc]{$P_{x_k\mid
      I_k}$} \psfrag{u_k}[cc][cc]{$u_k$} \psfrag{z_k}[cc][cc]{$z_k$}
  \psfrag{u_k-1}[cc][cc]{$u_{k-1}$}
  \psfrag{P_k-1}[cc][cc]{$P_{x_{k-1}\mid I_{k-1}}$}
  \includegraphics{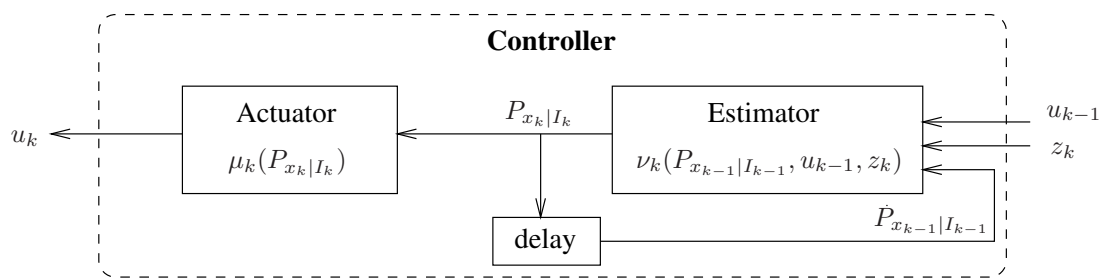}
\end{center}
\caption{On-Line Controller Architecture With Recursive Estimation} 
\label{fig:genArchRec}
\end{figure}

The function $Q_k(P_{x_k|I_k},u_k)$ has a very appealing interpretation;
namely, it represents the optimal cost-to-go at stage $k$ given the
probabilistic state and assuming the control $u_k$ will be selected.
Thus, it can be referred to as a \underline{control-dependent}
expected cost-to-go and, therefore, the minimization operation with
respect to control maps to $J_k$. This decoupling of the future costs
from the previous data is the primary appeal of a recursive estimation
scheme and goes hand-in-hand with performing the off-line strategy
optimization over the (non-expanding dimensional) sufficient
statistic. Recall that without recursive estimation, this decoupling
was apparent in the strategy implied by the cost-to-go calculation but
not in the cost-to-go calculation itself.  Figure~\ref{fig:genArchRec}
summarizes the on-line implementation of the optimal controller when
endowed with recursive estimation. The main advantage compared to the
implementation depicted in Fig.~\ref{fig:genArchSep} is the reduced
memory requirements as $k$ increases, and the reduced dimensionality
of the sufficient statistic as compared to the information vector also
implies an on-line computational advantage in the estimator module.

As a final note, the probabilistic state $P_{x_k\mid I_k}$ is not
necessarily the only sufficient statistic that can be found for a
specified system and measurement model. For instance, any other
expression that can uniquely determine the full probabilistic state
distribution will also be a sufficient statistic, and thus the points
of the above discussion remain valid. The next two sections provide
two specific examples where the specialized forms of the models yield
very simply parameterized probabilistic states. In the case of a
PO-MDP model (see section~\ref{POMDP}), the state space is a finite set and
thus the probabilistic state is simply a probability mass function
that can be characterized by a finite-dimensional vector with
non-negative values that sum to unity. In the case of the LQGR model
(see section~\ref{LQGR}), the probabilistic state is known to be a
multi-variate Gaussian distribution and thus a mean vector and
covariance matrix suffice.

\section{Partially-Observable Markov Decision Process \label{POMDP}}
{\it Markov Decision Process} (MDP) models describe a particular class
of multi-stage stochastic control problems that have been studied
extensively and repeatedly applied within, for example, the realm of
operations research, economics, computer and communication networks \cite{BeT02:IntPr, Gal96:DisSP}.
The imperfect state information counterpart to an MDP model is called
a {\it Partially-Observable} (PO)-MDP. Dynamic programming techniques
are especially well-developed for PO-MDP models with finite state,
control and measurement spaces.  In these cases, the probabilistic
state is simply a probability mass function over the original finite
state space, which can be expressed as a finite-dimensional vector
whose non-negative elements sum to unity. The estimation strategy can
be expressed in closed-form for the given system and measurement
models. The dynamic programming recursions that characterize the
actuation strategy can be expressed concisely using elementary matrix
operations that are easily implemented via computer.

\subsection{Standard Formulation/Results}

\subsubsection{Mathematical Models}
The specification of a PO-MDP begins with that of a regular MDP, with
the $k$th stage usually defined in terms of an underlying controlled
Markov chain. Letting $n_k$ and $m_k$ denote the (finite) number of
states and controls, respectively, the $k$th stage of an MDP is most
often specified by a set of $m_k\cdot n_k\cdot n_{k+1}$ {\it
  transition probabilities} $p(x_{k+1}\mid x_k,u_k)$ and {\it
  transition costs} $c(x_k,u_k,x_{k+1})$. Note that there is no need to
explicitly consider a terminal cost since it can be absorbed
into the specification of the transition costs for stage $K-1$. In
addition to the definition of the underlying MDP, the $k$th stage of a
PO-MDP involves one of $s_k$ possible measurements generated (for
stages $k > 0$) according to $m_{k-1} \cdot n_k\cdot s_k$ {\it
  observation probabilities} $q(z_k\mid x_k,u_{k-1})$. The initial
observation at $k = 0$, before any control has been applied, is
characterized by $n_0 \cdot s_0$ observation probabilities $q(z_0\mid
x_0)$.

In terms of the notation summarized in Table~\ref{ModParams}, the $k$th
stage of a PO-MDP is structurally characterized by the finite spaces
\begin{eqnarray*}
S_k = \{1,2,\ldots,n_k\}, \:\:
C_k = \{1,2,\ldots,m_k\}, \:\: \mbox{and} \:\:
Z_k = \{1,2,\ldots,s_k\} \qquad .
\end{eqnarray*}
The system equation can be expressed simply as
\begin{eqnarray*}
x_{k+1} = f_k(x_k,u_k,w_k) = w_k
\end{eqnarray*}
where the disturbance space $W_k = S_{k+1}$ and the disturbance
distribution $P_{w_k\mid x_k,u_k}$ corresponds directly to the
state-transition probabilities. Note that the conditional dependence
of the transition probabilities on the selected control $u_k$ reflects
the extent to which a controller can influence the subsequent
evolution of the state. Because $x_{k+1} = w_k$, the cost equation
is given directly by the transition costs
\begin{eqnarray*}
g_k(x_k,u_k,w_k) = c(x_k,u_k,x_{k+1}) \qquad .
\end{eqnarray*}
Similar to the system equation, for $k > 0$ the measurement equation can be
expressed as
\begin{eqnarray*}
z_k = h_k(x_k,u_{k-1},v_k) = v_k
\end{eqnarray*}
where the noise space $V_k = Z_k$ and the noise
distribution $P_{v_k\mid x_k,{\cal I}_{k-1}}$ corresponds directly to
the observation probabilities after recognizing that the set ${\cal
  I}_{k-1} = u_{k-1}$. Note that the conditional dependence of the
observation probabilities on the preceding control $u_{k-1}$ reflects
the extent to which the opportunity for sensor management can be
modeled. The measurement equation for $k = 0$ is analogously defined except
for the lack of dependence on any preceding control.

All of the stage-$k$ parameters of a PO-MDP can be conveniently
organized into a family of control-dependent matrices: $m_k$
transition probability matrices ${\bf F}_k$ and $m_k$ transition cost
matrices ${\bf G}_k$, all with dimension $n_k\times n_{k+1}$, as well
as $m_{k-1}$ observation probability matrices ${\bf H}_k$ with
dimension $n_k\times s_k$. More precisely, for each $u \in C_k$, $i
\in S_k$ and $j \in S_{k+1}$, let the specified transition probability
$p(j\mid i,u)$ correspond to the element in the $i$th row and $j$th
column of matrix ${\bf F}_k(u)$. Similarly, let the specified
transitions costs $c(i,u,j)$ populate the matrices ${\bf G}_k(u)$.
Finally, for each $u \in C_{k-1}$, $i \in S_k$ and $o \in Z_k$, let
the specified observation probability $q(o\mid i,u)$ correspond to the
element in the $i$th row and $o$th column of matrix ${\bf H}_k(u)$.
For $k = 0$, we simply have the single $n_0 \times s_0$
observation matrix ${\bf H}_0$ populated by the initial observation
probabilities $q(z_0\mid x_0)$.

\subsubsection{Estimation Strategy}
Because each stage $k$ of a PO-MDP involves a discrete and finite
state space $S_k$ with cardinality $n_k$, the probabilistic state
$P_{x_k\mid I_k}$ has the form of a standard probability mass function
(PMF). We represent this PMF as a column vector ${\bf p}_k$ of length
$n_k$, where the $i$th component corresponds to the likelihood that
the current state $x_k$ is the $i$th element of space $S_k$. Note
that, if state $i$ is without a doubt the true state, then ${\bf p}_k
= 1_k(i)$ corresponds to the unit vector in the $i$th dimension;
conversely, when no meaningful information has been provided to the
estimator, ${\bf p}_k$ might be the uniform distribution.

A straightforward application of Bayes' rule, combined with the
compactness of the transition and observation probability matrices, then
yields the estimation strategy 
\begin{eqnarray}
\begin{array}{lcl}
{\bf p}_0 & = & \nu_0(I_0) = P_{x_0\mid z_0} = \displaystyle 
\frac{\left[{\bf H}_0\right]_{z_0} * P_{x_0}} 
{\left(\left[{\bf H}_0\right]_{z_0}\right)' P_{x_0}} \\[0.4in]
{\bf p}_k & = & \nu_k({\bf p}_{k-1}, u_{k-1}, z_k) = \displaystyle 
\frac{\left[{\bf H}_k(u_{k-1})\right]_{z_k} * 
\left( {\bf F}_{k-1}(u_{k-1})'{\bf
p}_{k-1}\right)}{\left(\left[{\bf H}_k(u_{k-1})\right]_{z_k}\right)' {\bf
F}_{k-1}(u_{k-1})'{\bf p}_{k-1}} \: ,  \quad  k = 1, 2, \ldots, K-1 
\end{array}
\label{eq:POMDPest}
\end{eqnarray}
where ${\bf A}'$ denotes the transpose of matrix ${\bf A}$,
$\left[{\bf A}\right]_o$ is the column vector corresponding to the
$o$th column of matrix ${\bf A}$, and the usual matrix multiplications
are implied except that ${\bf A}_1 * {\bf A}_2$ denotes the
\underline{element-wise} multiplication of two matrices. The product
${\bf F}_{k-1}(u_{k-1})'{\bf p}_{k-1}$ represents a {\it prediction
  step} of the estimator accounting for the control-dependent dynamics
as reflected by the state transition probabilities; the operation
involving $[{\bf H}_k(u_{k-1})]_{z_k}$ represents a {\it correction
  step} of the estimator once the measurement $z_k$ is observed. Note
also that the computation in the denominator serves merely to
normalize the vector resulting from the computation in the numerator.

\subsubsection{Actuation Strategy}
Due to the finite cardinality of the control space, the function
$Q_k(P_{x_k\mid I_k},u_k)$ can be defined as a length-$m_k$ vector
${\bf q}_k({\bf p}_k)$ associated to any given probabilistic state
vector ${\bf p}_k$. Using the same matrix notation as in
(\ref{eq:POMDPest}), the real-valued element $[{\bf q}_k({\bf
  p}_k)^\prime ]_u$ corresponds to the expected cost-to-go at stage
$k$ given probabilistic state vector ${\bf p}_k$ and assuming selected
control $u_k = u$. The cost-to-go function $J_k$ and associated
optimal actuation policy $\mu_k^*$ are given by
\begin{eqnarray}
J_k({\bf p}_k) = \min_{u \in \{1,2,\ldots,m_k\}}
 \left[{\bf q}_k({\bf p}_k)^\prime\right]_u \qquad \Rightarrow \quad 
\mu_k^*({\bf p}_k) = \arg \min_{u \in \{1,2,\ldots,m_k\}}
 \left[{\bf q}_k({\bf p}_k)^\prime\right]_u \qquad .
\label{eq:POMDPactJ} 
\end{eqnarray}
We can also apply the compact matrix notation to the iteration that,
combined with (\ref{eq:POMDPactJ}), computes the vectors ${\bf
  q}_k({\bf p}_k)$.  Letting ${\bf e_\ell}$ denote the length-$\ell$
column vector of all 1's, we have for every $u \in \{1, 2, \ldots,
m_k\}$
\begin{eqnarray}
\begin{array}{rcl}
[{\bf q}_{K-1}({\bf p}_{K-1})^\prime]_u & = & {\bf p}_{K-1}^\prime 
\left( {\bf F}_{K-1}(u) * {\bf G}_{K-1}(u) \right) {\bf e}_{n_K} \\[0.2in]
[{\bf q}_k({\bf p}_k)^\prime]_u & = & {\bf p}_k^\prime
\left( {\bf F}_k(u) * {\bf G}_k(u) \right) {\bf e}_{n_{k+1}} + \\[0.1in] 
& & \qquad \qquad \qquad \qquad 
\displaystyle \sum_{z = 1}^{s_{k+1}} {\bf p}_k^\prime 
\left( {\bf F}_k(u)\left[{\bf H}_{k+1}(u)\right]_z 
\right) J_{k+1}\left(\nu_{k+1}\left({\bf p}_k,u,z\right)\right)
\end{array}
\label{eq:POMDPactQ}
\end{eqnarray}

The dynamic programming algorithm characterized by
(\ref{eq:POMDPest})-(\ref{eq:POMDPactQ}) admits mathematical analysis
that demonstrates many desirable properties. It turns out that the
cost-to-go functions $J_k$ are {\it piecewise linear} and {\it
  concave} functions of ${\bf p}_k$. This implies that each $J_k$ can
be characterized by a finite set of scalar values corresponding to the
breakpoints and slopes of the constant line segments.  However, for a
fixed $k$, as the horizon $K$ increases the (finite) number of such
scalars grows exponentially, in general. On the bright side, for any
element ${\bf p}_k$ out of the space of all length-$n_k$ probabilistic
state vectors, these scalars ultimately reduce to the finite
parametrization defining the vector ${\bf q}_k({\bf p}_k)$, and thus
an actuation strategy for a PO-MDP is quite amenable to various
approximation methods. That is, finding the optimal actuation policy
at each stage $k$ reduces to a search for a specific function ${\bf
  q}_k$, mapping the space of all length-$n_k$ vectors whose
nonnegative elements sum to unity to the $m_k$-dimensional Euclidean
space.

\subsection{Illustrative Example (Machine Repair with Diagnosis Option) }
A machine can be in one of two states at the beginning of any time
period: the good state corresponds to the machine in proper condition
and the bad state to the machine in improper condition.  After one
time period of operation, it stays in the good or the bad state with
probability $(1-\alpha)$ or probability 1, respectively.  Just after
the beginning of each time period, an optional repair action can be
taken; if taken, a cost $c_R>0$ is incurred but the machine is
guaranteed to operate in the good state during that time period.  In
the absence of the repair action, a single time period of operation in
the good state is cost-free but a time period of operation in the bad
state incurs cost $c_B>c_R$.

The true state in each time period is not known but, just prior to the
moment that the optional repair action can be taken, a sensor outputs
one of two messages regarding the condition of the machine.  However,
the sensor is not completely reliable; it is known to incorrectly
indicate the bad state with probability $\eta_f$ (i.e., the false alarm
probability) and correctly indicate the bad state with probability
$\eta_d$ (i.e., the true detection probability). It is also possible to
supplement the sensor with the output of a diagnostic test that must be
initiated in the previous time period, incurring an additional cost
$c_D \geq 0$. The combined measurement, while perhaps costly, is perhaps
also more accurate, yielding a false alarm probability of $\gamma_f
\leq \eta_f$ and a true detection probability $\gamma_d \geq
\eta_d$.

Assume it is not possible to both repair the machine and initiate a
diagnostic test in the same time period. Also assume that, before any
measurements are received and any actions are taken, that the machine
is already in the bad state with probability $\alpha$. The terminal cost
is zero.

\subsubsection{The Model}
It is easy to see that the above problem fits into the general PO-MDP
model. In fact, we have the special case of a {\it stationary} PO-MDP
model, meaning the model parameters do not depend on stage $k$. For
all $k$, the finite state, control and measurement spaces are defined
by $n_k = 2$, $m_k = 3$ and $s_k = 2$, respectively. The family of
control-dependent matrices are given by
\begin{eqnarray}
{\bf F}_k(u_k) & = & \left\{ 
\begin{array}{ccl}
\left[ \begin{array}{cc}
1-\alpha & \alpha \\
0 & 1
\end{array} \right] & , & u_k = 1,2 \\[0.25in]
\left[ \begin{array}{cc}
1-\alpha & \alpha \\
1-\alpha & \alpha
\end{array} \right] & , & u_k = 3 \\
\end{array}
\right. \quad \mbox{for $k = 0,1,\ldots,K-1$} \qquad , 
\label{eq:mrSys} \\[0.25in]
{\bf G}_k(u_k) & = & \left\{ 
\begin{array}{ccl}
\left[ \begin{array}{cc}
0 & 0 \\
c_B & c_B
\end{array} \right] & , & u_k = 1 \\[0.25in]
\left[ \begin{array}{cc}
c_D & c_D \\
c_D + c_B & c_D + c_B
\end{array} \right] & , & u_k = 2 \\[0.25in]
\left[ \begin{array}{cc}
c_R & c_R \\
c_R & c_R
\end{array} \right] & , & u_k = 3
\end{array}
\right. \quad \mbox{for $k = 0,1,\ldots,K-1$} \qquad \mbox{and} \qquad 
\label{eq:mrCos} \\[0.25in]
{\bf H}_k(u_{k-1}) & = & \left\{ 
\begin{array}{ccl}
\left[ \begin{array}{cc}
1-\eta_f & \eta_f \\
1-\eta_d & \eta_d
\end{array} \right] & , & u_{k-1} = 1,3 \\[0.25in]
\left[ \begin{array}{cc}
1-\gamma_f & \gamma_f \\
1-\gamma_d & \gamma_d
\end{array} \right] & , & u_{k-1} = 2 \\
\end{array}
\right. \quad \mbox{for $k = 1, 2, \ldots, K-1$}
\label{eq:mrMea}
\end{eqnarray}
where states 1 and 2 represent the good and bad state; controls 1,2
and 3 represent the default, diagnose and repair action; and
measurements 1 and 2 represent the probably-good and probably-bad
measurement output. Note also that the initial measurement model is
simply ${\bf H}_0 = {\bf H}_k(1)$ and the initial state distribution
is given by
\begin{equation}
P_{x_0} = \left[ \begin{array}{c} 1-\alpha \\ \alpha \end{array} \right]
\qquad .
\label{eq:mrIni}
\end{equation}

Intuition might lead to a conjecture that the behavior of the optimal
strategy should, at each stage $k$, select the default action when our
confidence in the bad state is lowest (i.e., $\rho_k \rightarrow 0$),
select the repair action when this confidence is highest (i.e., $\rho_k
\rightarrow 1$) and select the a diagnostic test when this confidence is
least polarized toward either machine state (i.e., $\rho_k \rightarrow
0.5$).  Certainly the thresholds differentiating these decision regions
will vary at each stage, but perhaps the parameterized structure of the
optimal actuation policy is constant. In the analysis and computational
examples that follow, we provide evidence that this is probably true,
but some subtleties arise concerning the relative positions of these
thresholds. In some cases, depending on the numerical values of the
model parameters, these threshold positions yield an optimal strategy 
with the \underline{appearance} of having an appreciably different structure.

\subsubsection{The Estimation Strategy}
The probabilistic state is the PMF ${\bf p}_k$ that, at each stage $k$
in this machine repair problem, is a length-2 column vector whose
elements sum to unity. Thus, we can define a scalar
$$
\rho_k = \Pr(x_k=2|I_k) \qquad \Rightarrow \qquad 
{\bf p}_k = \left[ \begin{array}{c} 1-\rho_k \\ \rho_k \end{array} \right]
$$ and let it completely parameterize the probabilistic state
vector. Substitution of (\ref{eq:mrSys}), (\ref{eq:mrMea}) and
(\ref{eq:mrIni}) into (\ref{eq:POMDPest}) yields, after some algebra, an
estimation strategy of
\begin{equation}
\begin{array}{rcl}
\rho_0 & = & \left\{ \begin{array}{ccl} {\displaystyle
\frac{(1-\eta_d)\alpha}{(1-\eta_f)(1-\alpha) + (1-\eta_d)\alpha}} & , &
z_0 = 1 \\[0.25in] {\displaystyle \frac{\eta_d \alpha}{\eta_f(1-\alpha)
+ \eta_d \alpha}} & , & z_0 = 2
\end{array} \right. \qquad \mbox{and, for $k = 1, 2, \ldots, K-1$,} \\[0.75in]
\rho_k & = & \left\{ \begin{array}{ccl} 
{\displaystyle \frac{(1-\eta_d)\left[ \alpha + (1-\alpha)\rho_{k-1}\right]}
{(1-\eta_f)(1-\alpha )(1-\rho_{k-1}) + 
(1-\eta_d)\left[ \alpha + (1-\alpha)\rho_{k-1}\right]}} 
& , & (u_{k-1},z_k) = (1,1) \\[0.25in]
{\displaystyle \frac{\eta_d \left[ \alpha + (1-\alpha)\rho_{k-1}\right]}
{\eta_f (1-\alpha )(1-\rho_{k-1}) + 
\eta_d \left[ \alpha + (1-\alpha)\rho_{k-1}\right]}} 
& , & (u_{k-1},z_k) = (1,2) \\[0.25in]
{\displaystyle \frac{(1-\gamma_d)\left[ \alpha + 
(1-\alpha)\rho_{k-1}\right]}{(1-\gamma_f)(1-\alpha )(1-\rho_{k-1}) + 
(1-\gamma_d)\left[ \alpha + (1-\alpha)\rho_{k-1}\right]}} 
& , & (u_{k-1},z_k) = (2,1) \\[0.25in]
{\displaystyle \frac{\gamma_d \left[ \alpha + (1-\alpha)\rho_{k-1}\right]}
{\gamma_f (1-\alpha )(1-\rho_{k-1}) + 
\gamma_d \left[ \alpha + (1-\alpha)\rho_{k-1}\right]}} 
& , & (u_{k-1},z_k) = (2,2) \\[0.25in]
{\displaystyle \frac{(1-\eta_d)\alpha}{(1-\eta_f)(1-\alpha ) + 
(1-\eta_d)\alpha}}
& , & (u_{k-1},z_k) = (3,1) \\[0.25in]
{\displaystyle \frac{\eta_d \alpha}{\eta_f (1-\alpha ) + 
\eta_d \alpha}}
& , & (u_{k-1},z_k) = (3,2) \\[0.25in]
\end{array} \right. 
\end{array}
\label{eq:mrEst}
\end{equation}
expressed with respect to a recursion on parameter $\rho_k$.

\subsubsection{The Actuation Strategy}
Given the estimation strategy in (\ref{eq:mrEst}), where the
probabilistic state at each stage $k$ is conveniently parameterized by
the scalar $\rho_k$, we now substitute (\ref{eq:mrSys})-(\ref{eq:mrMea})
into the algorithm characterized by (\ref{eq:POMDPactJ}) and
(\ref{eq:POMDPactQ}). Beginning at stage $K-1$, we first obtain
$$
[{\bf q}_{K-1}({\bf p}_{K-1})^\prime]_1 = 
\left[ \begin{array}{c} 1 - \rho_{K-1} \\ \rho_{K-1} \end{array} 
\right]^\prime \left( \left[ \begin{array}{cc} 1-\alpha &
  \alpha \\ 0 & 1 \end{array} \right] * \left[ \begin{array}{cc} 0 & 0
  \\ c_B & c_B \end{array} \right] \right) \left[ \begin{array}{c} 1 \\ 1
\end{array} \right] = \left[ \begin{array}{c} 0 \\ \rho_{K-1} c_B 
\end{array} \right]
$$
and, repeating this calculation for $u = 2$ and $u = 3$, we can
form the vector
$$
{\bf q}_{K-1}({\bf p}_{K-1}) = \left[ \begin{array}{c} 
\rho_{K-1} c_B \\ \rho_{K-1} (c_D + c_B) \\ c_R
\end{array} \right] \qquad .
$$ 
Thus, it follows that the optimal cost-to-go from stage $K-1$ is given
by
\begin{equation}
J_{K-1}({\bf p}_{K-1}) = \left\{ \begin{array}{ccl} 
\rho_{K-1}c_B & , & \rho_{K-1} \leq c_R / c_B \\
c_R & , & \rho_{K-1} > c_R / c_B \\
\end{array} \right. ,
\label{eq:mrActJ}
\end{equation}
where we have favored the default action if the minimization involves
a tie; notice the piecewise-linear and concave dependence on the
probabilistic state. The corresponding optimal actuation policy is
then
$$
\mu^*_{K-1}({\bf p}_{K-1}) = \left\{ \begin{array}{ccl} 
1 & , & \rho_{K-1} \leq c_R / c_B \\
3 & , & \rho_{K-1} > c_R / c_B
\end{array} \right. \qquad .
$$
It is no surprise, at this final stage, that we will never select
the costly diagnostic test---after all, the modeled decision process
terminates before receiving a next measurement.

We now proceed to the next iteration of the dynamic programming
algorithm, which begins with the computation of the vector ${\bf
  q}_{K-2}({\bf p}_{K-2})$ via (\ref{eq:POMDPactQ}).  For each $u$,
the first term reflects the immediate cost and is the same calculation
as performed in stage $K-1$ e.g., with $u=1$,
\[
{\bf p}_k^\prime \left( {\bf F}_k(1) * {\bf G}_k(1) \right) {\bf e}_2 = 
\left[ \begin{array}{c} 0 \\ \rho_{K-2} c_B \end{array} \right] \quad .
\]
However, calculation of the second term, reflecting the expected
future cost, involves several more steps. Assume vector ${\bf p}_k$ is
given and consider the calculation for $u=1$. We must first evaluate
(\ref{eq:mrActJ}) at each particular probabilistic state resulting from
(\ref{eq:mrEst}) for each of the $2$ possible measurements $z_{K-1}$.
Appropriately weighting these terms and summing together with the
immediate cost yields the real-valued element $[{\bf q}_{K-2}({\bf
  p}_{K-2})^\prime]_1$.  Repeating for $u = 2$ and $u=3$ yields the
vector ${\bf q}_{K-2}({\bf p}_{K-2})$ and, in turn, enables the
computation of cost-to-go function $J_{K-2}$ and the implied optimal
policy $\mu^*_{K-2}$ according to (\ref{eq:POMDPactJ}). We would
similarly proceed with the next iteration of the dynamic programming
algorithm.

Even for this simple example, it is clear that an analytical solution
for more than $K=2$ stages of the dynamic programming algorithm becomes
extremely cumbersome; indeed, this is an unavoidable manifestation of
the aforementioned exponential complexity inherent to the exact
algorithm.  By selecting numerical values for all involved model
parameters, we can rely on a computer to perform these cumbersome
calculations and readily compute the solution for slightly larger values
of $K$. Of course, as we continue to increase the number of stages $K$,
the same complexity issue will eventually also overwhelm such a
computational approach. Fortunately, for moderate values of $K$ where a
computational solution is feasible in this example, we can still observe
some interesting behavior of the optimal solution.

\begin{figure}[t!]
\begin{center}
{\bf Parameters:} $(\alpha,\eta_f,\eta_d,\gamma_f,\gamma_d) = 
(0.2,0.3,0.7,0.1,0.9)$, \quad $(c_R,c_B,c_D) = (5,10,1)$ \\[0.3in]
\resizebox{0.75\textwidth}{!}{\includegraphics{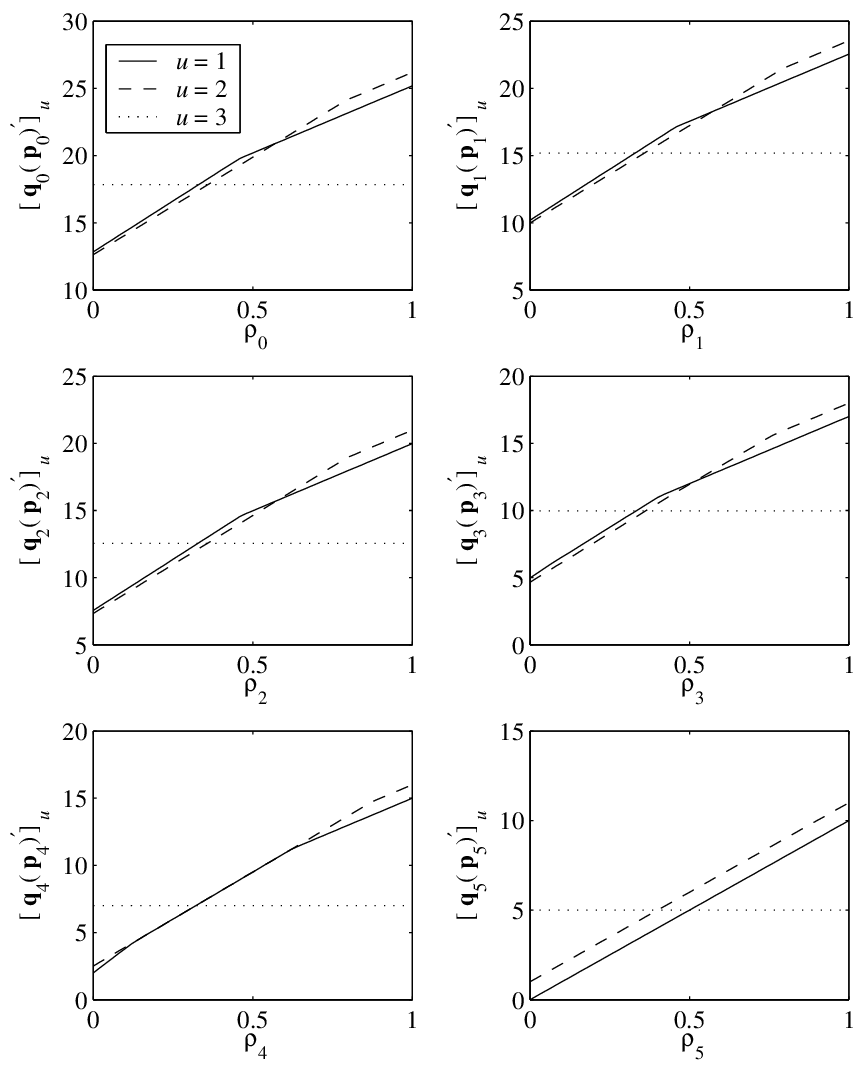}}
\caption{An Optimal Actuation Strategy for the Machine Repair Example
($K=6$)}
\label{fig:mrEx1}
\end{center}
\end{figure}

Let us first assume numerical values
\begin{eqnarray*}
\mbox{\bf Probabilities:} & &
(\alpha,\eta_f,\eta_d,\gamma_f,\gamma_d) = 
(0.2,0.3,0.7,0.1,0.9) \\
\mbox{\bf Costs:} & & (c_R,c_B,c_D) = (5,10,1) 
\end{eqnarray*}
and then consider $K=6$ total decision stages. Figure~\ref{fig:mrEx1}
illustrates the \underline{control-dependent} cost-to-go function, or
the vector components $[{\bf q}_k({\bf p}_k)^\prime]_u$ $(u=1,2,3)$
versus scalar parameter $\rho_k$, at each stage $k = 0, 1,\ldots,5$.
Recall that the minimization of this quantity for each $k$ defines the
optimal cost-to-go function $J_k$; thus, for any value of parameter
$\rho_k$, the optimal cost-to-go corresponds to the ``lowest'' of the
three piecewise-linear curves that, in turn, corresponds to the optimal
control $u_k^*$.  In the plot of ${\bf q}_5({\bf p}_5)$, note that the
diagnostic test is never favored, agreeing with the conclusion reached
in the analytical solution for the last stage $k=5$. Also in agreement
with the analytical solution, when the level-of-confidence in the bad
state exceeds a certain threshold, the repair action is taken and
otherwise the default action is taken.  The policy in the second-to-last
stage $k=4$ also never favors the diagnostic action, although for
moderate values of $\rho_k$ it nearly does so. Even in the
second-to-last stage, the benefit of a more accurate measurement does
not outweigh the additional cost to obtain the increased accuracy.
In the earlier stages, however, provided we do not take the repair
action, we always favor the more accurate measurement.

\begin{figure}[t!]
\begin{center}
{\bf Parameters:} $(\alpha,\eta_f,\eta_d,\gamma_f,\gamma_d) = 
(0.02,0.3,0.7,0.1,0.9)$, \quad $(c_R,c_B,c_D) = (5,10,1)$ \\[0.3in]
\resizebox{0.75\textwidth}{!}{\includegraphics{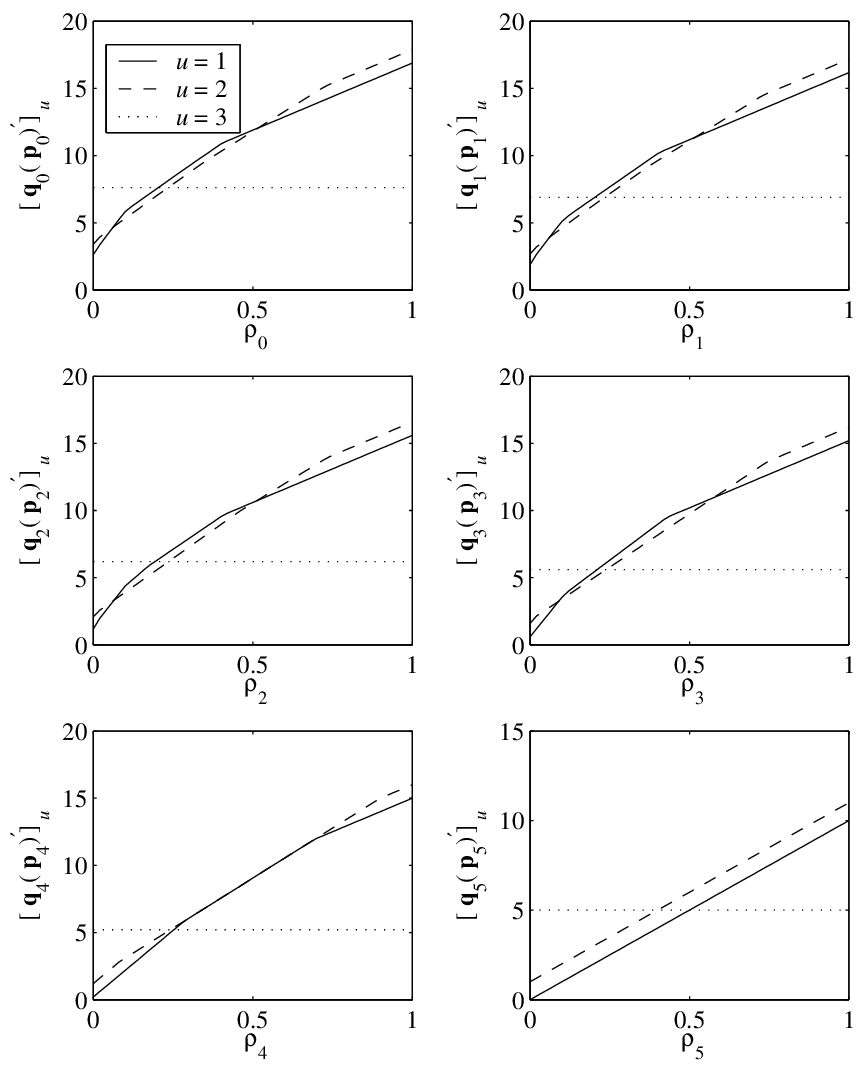}}
\caption{Optimal Actuation Strategy with Increased Machine Failure Rate
($K=6$)}
\label{fig:mrEx2}
\end{center}
\end{figure}

\begin{figure}[t!]
\begin{center}
{\bf Parameters:} $(\alpha,\eta_f,\eta_d,\gamma_f,\gamma_d) = 
(0.35,0.3,0.7,0.1,0.9)$, \quad $(c_R,c_B,c_D) = (5,50,1)$ \\[0.3in]
\resizebox{0.75\textwidth}{!}{\includegraphics{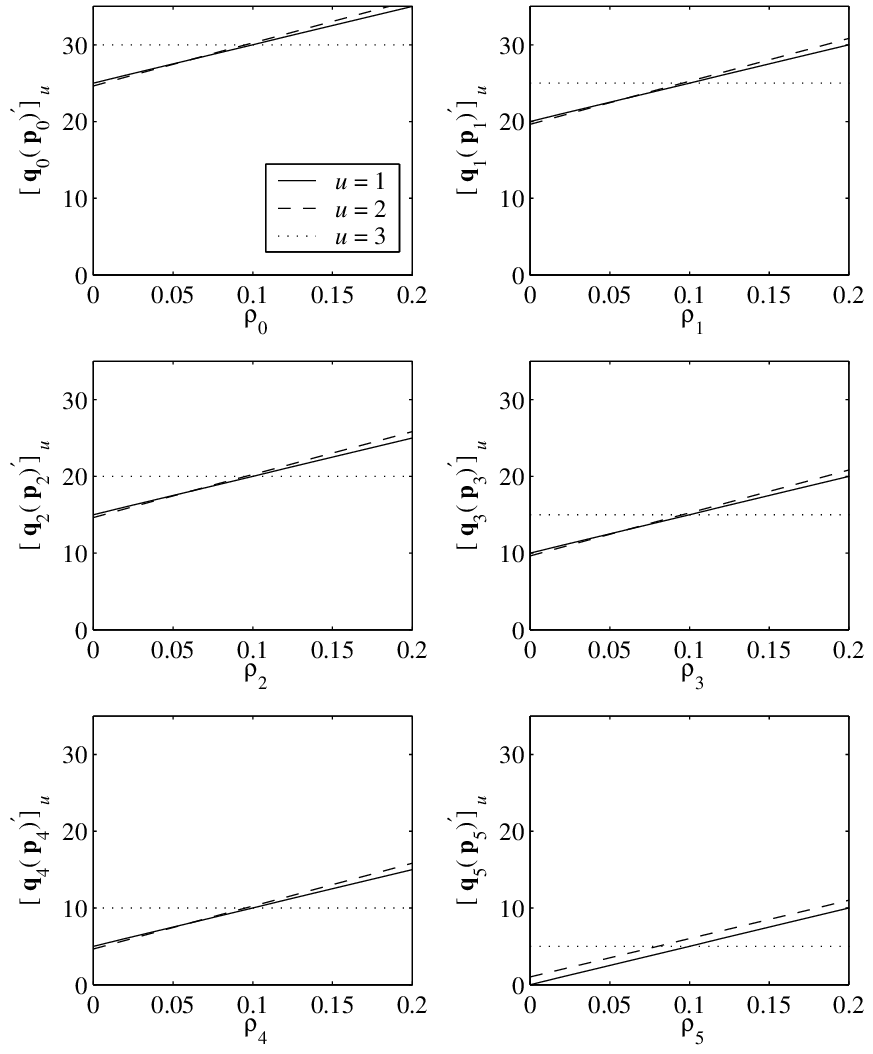}}
\caption{Optimal Actuation Strategy with Increased Improper-Operation Cost
($K=6$)}
\label{fig:mrEx3}
\end{center}
\end{figure}

The actuation strategy depicted in Fig.~\ref{fig:mrEx1} arguably agrees
with the aforementioned intuition; the lowest values of $\rho_k$ map to
the default action, more moderate values to the diagnostic action and
highest values to the repair action. In the earlier stages, it just so
happens that the threshold between the default and diagnosis actions is
below zero, the lower bound on parameter $\rho_k$. This is exhibited
more clearly in Fig.~\ref{fig:mrEx2}, which is the same data as in
Fig.~\ref{fig:mrEx1} except that parameter $\alpha = 0.02$.
Note how the piecewise-linear curves corresponding to the default and
diagnostic actions ($u=1$ and $u=2$, respectively) compare over the
range of parameter $\rho_k$. As intuition suggests, the diagnostic
action is favored for moderate values of $\rho_k$ but, at either extreme
of decreased uncertainty in the true state, the default action is
favored. The position of the horizontal line corresponding to the repair
action relative to the two other curves can yield apparently different
policy structure. Figure~\ref{fig:mrEx3} shows the optimal actuation
strategy (zoomed in at values of $\rho_k$ of interest) when parameters
$\alpha$ and $c_B$ are changed to $0.35$ and $50$, respectively. Here,
lowest values of $\rho_k$ map to the diagnostic action and moderate
values of $\rho_k$ map to the default action, appearing to be a
different structure than what our intuition suggests. But, as in the
earlier stages in Fig.~\ref{fig:mrEx1}, the lower threshold between the
default and diagnostic curves is below zero; moreover, the upper
threshold between the same curves is below the threshold at which the
horizontal repair line intersects the other curves. This instance of the
policy reflects the exceedingly high failure cost combined with the
relatively high single-stage machine failure rate.

It is worth contemplating the implications of the previous
figures. While, in this example, intuition reliably identified some
interesting properties of the optimal strategy, its specific realization
is heavily dependent on the model parameters.  Indeed, understanding the
impact on the control strategy and the various model parameters is
fundamental for stochastic control problems, including especially
problems of sensor management whose benefit/cost ratios inherently
extend over the temporal dimension! 

\begin{figure}[t!]
\begin{center}
{\bf Parameters:} $(\alpha,\eta_f,\eta_d,\gamma_f,\gamma_d) =
(0.2,0.3,0.7,0.1,0.9)$, \quad $(c_R,c_B,c_D) = (5,10,1)$ \\[0.3in]
\resizebox{0.75\textwidth}{!}{\includegraphics{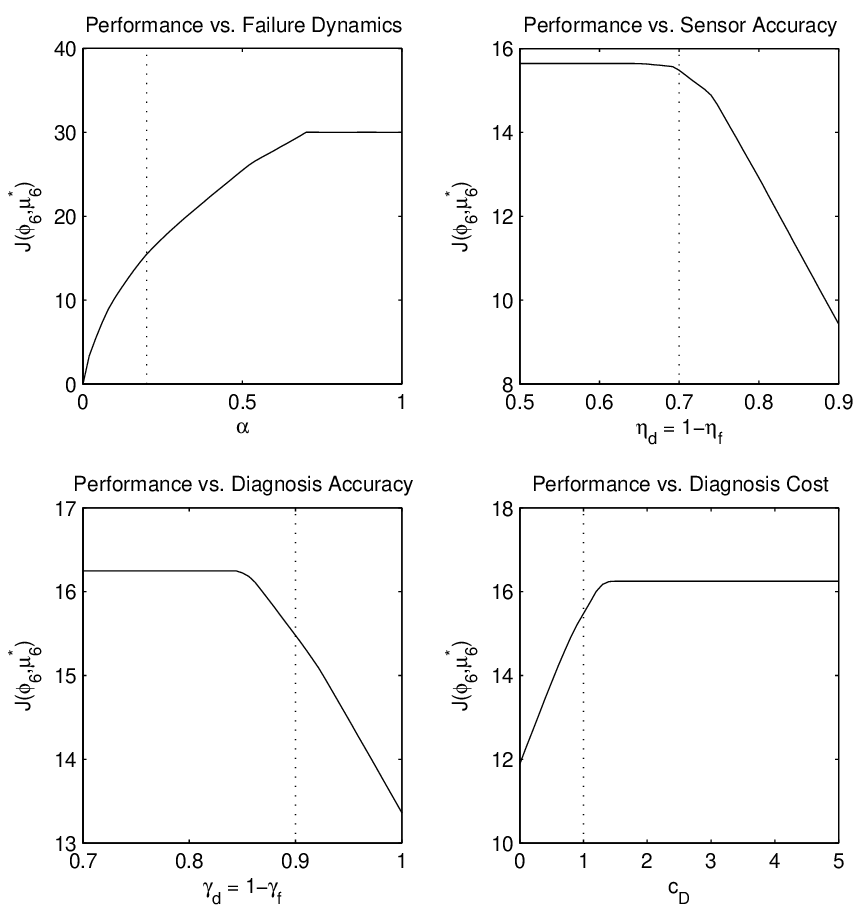}}
\caption{Sensitivity of Optimal Total Cost in Machine Repair Example ($K=6$)}
\label{fig:mrEx4}
\end{center}
\end{figure}

We close the discussion of our machine repair example with a
sensitivity analysis of the optimal control performance, or the
expected total cost $J(\phi_K,\mu^*_K|\bullet)$ defined in
(\ref{eq:CLCostRec}), when varying (i) state dynamics captured by
parameter $\alpha$, (ii) sensor accuracy captured by parameters
$\eta_f$ and $\eta_d$ (we constrain $\eta_d = 1-\eta_f$ and $\eta_d
\leq \gamma_d$ here), (iii) combined sensor/diagnosis accuracy
captured by parameters $\gamma_f$ and $\gamma_d$ (we constrain
$\gamma_d = 1-\gamma_f$ and $\gamma_d \geq \eta_d$ here) and (iv) the
cost associated to obtaining a more accurate measurement captured by
parameter $c_D$ (we constrain $c_D \leq c_R$. Each plot in
Fig.~\ref{fig:mrEx4} shows the respective sensitivity curve, where the
parameters have nominal values corresponding to those used to generate
Fig.~\ref{fig:mrEx1}. The vertical dashed line in each plot indicates
the nominal value of each respective varying parameter. The following
conclusions about the machine repair example, assuming the specified
nominal parameter values, can be drawn from these plots: (i) as the
single-stage failure rate $\alpha$ of the machine increases, so does
the total cost, plateauing at a cost of $30 = Kc_R$ at which point the
optimal strategy is simply to repair at every stage; (ii) increased
sensor accuracy leads to improved performance, and a sensor that is
incorrect more than $30\% $ of the time is practically worthless
without the diagnostic action; (iii) a unit-cost diagnostic action
appears to only be of benefit if it improves measurement accuracy by
at least $15\%$; and, (iv) a 20\% improvement in measurement accuracy
is only worthwhile if it costs less than 1.2 (approximately).

\section{Linear-Quadratic-Gaussian Regulator \label{LQGR}}
{\it Linear-Quadratic} (LQ) models describe a particular class of
multi-stage control problems where the objective is to maintain the
state of a stochastic process as well as the selected controls close
to specified target trajectories. These target trajectories need not
satisfy the constraints implied by the system model, and thus the
optimal control strategy must balance between costs due to state
deviations with costs due to control deviations. A quadratic function
of the state and control signals is one way to induce higher penalties
for larger deviations from the target trajectories. When the system
model is also a sequence of linear equations, then the LQ problem
admits a nice closed-form analytical solution where the optimal
control at each stage is a linear function of the state. By relying on
the machinery of linear algebra \cite{Str93:LinAl}, this solution readily extends to
multi-input/multi-output systems and thus spans a broad spectrum of
practical applications. The LQ-Regulator (LQR) problem is the special
case where the target trajectories correspond to the origin; thus, the
quadratic cost corresponds to balancing between the energy of the
state signals with the energy of the applied control signals.

The stochastic counterpart of the LQR problem, yet still assuming
perfect state information at each stage, corresponds to the presence of
a random initial state $x_0$ and a stochastic disturbance $w_k$ that
enters linearly in the system equation. To preserve linearity of the
system model, the disturbance distribution must be assumed to be
independent of both $x_k$ and $u_k$. Under this condition, the
stochastic LQR solution is known to satisfy {\it certainty equivalence},
meaning the optimal strategy corresponds to that of the deterministic
problem if we replaced $x_0$ and $w_k$ with their expected values
$E[x_0]$ and $E[w_k]$, respectively. The achieved total expected cost is
equal to the total cost of the deterministic counterpart plus an
additive sequence of the (weighted) second moments of each random
variable.  Thus, the mathematics reveals the intuitive notion that
on-average control performance degrades as the level of uncertainty in
the process grows.

The LQGR model is the imperfect state information counterpart of the
stochastic LQR problem where the measurement model is also assumed to be
linear and the disturbance, initial state and noise distributions are
all Gaussian random vectors.  To preserve the linearity of the
measurement model, it is typically assumed that the noise distribution
does not depend on $x_k$ nor on any element of the set ${\cal
I}_{k-1}$. The celebrated results for the LQGR problem with the
mentioned model assumptions \cite{Ber95:DPOC1} are: 
\begin{itemize}
\item the probabilistic state $P_{x_k\mid I_k}$ at each stage $k$ is
  known to be a multi-variate Gaussian distribution and thus it is completely
  characterized by its second-order statistics, the mean vector ${\bf
    m}_{k\mid k}$ and a covariance matrix ${\bf \Sigma}_{k\mid k}$;
\item the recursive estimation strategy is equal to the classical
  Kalman filtering solution, which involves coupled recursions for
  both ${\bf m}_{k\mid k}$ and ${\bf \Sigma}_{k\mid k}$; 
\item the actuation strategy is very similar to the perfect state
  information case, only it is a linear function of ${\bf m}_{k \mid k}$
  instead of the true state.
\end{itemize}
The achieved expected total cost is equal to the total cost for the
perfect state information counterpart plus an additive sequence of the
(weighted) estimation error as quantified by the matrices ${\bf
\Sigma}_{k\mid k}$.  Thus, the mathematics reveals the intuitive notion
that lower quality state estimation degrades on-average control
performance.

\subsection{Standard Formulation/Results}
The following section presents the equations that define the standard
LQGR model and the associated estimation/actuation strategy. As
highlighted above, these analytic results depend strongly on the
assumptions that all modeled distributions are Gaussian random vectors
and, furthermore, that both the disturbance/noise distributions are
restricted to be independent of the states and controls. But, we will
see that modeling the opportunity for sensor management corresponds to
having the noise distribution in the measurement model depend, in some
way, on the preceding control. Depending on the manner in which this
control-dependence is included, the implications on the optimal
strategy can be severe.

\subsubsection{Mathematical Models}
In terms of the notation summarized in Table~\ref{ModParams}, the $k$th
stage of a LQGR is structurally characterized by the vector spaces
\begin{eqnarray*}
S_k = W_k = \Re^n \: , \quad C_k = \Re^m \quad \mbox{and} \quad Z_k =
V_k = \Re^s
\end{eqnarray*}
implying that the states, controls and measurements are all real-valued
vectors (which we indicate in bold notation) of respective dimension.
The linear system equation is
\begin{eqnarray*}
{\bf x}_{k+1} = f_k({\bf x}_k, {\bf u}_k, {\bf w}_k) = {\bf A}_k {\bf
x}_k + {\bf B}_k {\bf u}_k + {\bf w}_k
\end{eqnarray*}
where matrices ${\bf A}_k$ and ${\bf B}_k$ are of appropriate
dimensions, the Gaussian initial state distribution has mean ${\bf
m}_{x_0}$ and covariance matrix ${\bf \Sigma}_{x_0}$, and the Gaussian
disturbance distribution has zero-mean and a specified covariance matrix
${\bf \Sigma}_{w_k}$. The quadratic cost equation is expressed as
\begin{eqnarray*}
g_k({\bf x}_k, {\bf u}_k, {\bf w}_k) & = & {\bf x}_k'{\bf T}_k{\bf x}_k +
{\bf u}_k'{\bf R}_k{\bf u}_k \\
g_K({\bf x}_K) & = & {\bf x}_K'{\bf T}_K{\bf x}_K
\end{eqnarray*}
where both matrices ${\bf T}_k$ and ${\bf R}_k$ are symmetric and of
appropriate dimensions. For the regulation problem to be well-posed, we
assume ${\bf T}_k$ is positive semi-definite (i.e., there is never reward
for deviating any component of the state from the origin) and ${\bf
R}_k$ is positive definite (i.e., there is always some penalty for a
non-zero component in the control signal). Similar to the system
equation, the measurement equation is
\begin{eqnarray*}
{\bf z}_0 & = & h_0({\bf x}_0, {\bf v}_0) = {\bf C}_0 {\bf x}_0 + {\bf v}_0 \\
{\bf z}_k & = & h_k({\bf x}_k, {\bf u}_{k-1}, {\bf v}_k) = {\bf C}_k {\bf
x}_k + {\bf D}_{k-1} {\bf u}_{k-1} + {\bf v}_k 
\end{eqnarray*}
where matrices ${\bf C}_k$ and ${\bf D}_{k-1}$ are of appropriate dimensions
and the Gaussian noise distribution has zero-mean and specified covariance
matrix ${\bf \Sigma}_{v_k}$. 

Note that, by linearity of the system model and measurement model, the
zero-mean assumption on both the disturbance and noise distributions
can be made without loss of generality. We have also implied that the random
vectors ${\bf x}_0$, ${\bf w}_0, {\bf w}_1, \ldots, {\bf w}_{K-1}, {\bf
v}_0, {\bf v}_0, \ldots, {\bf v}_{K-1}$ are mutually independent. The
general specification of the system model (e.g., see Table
\ref{ModParams}) already implies each vector ${\bf w}_k$ can depend at
most on ${\bf x}_k$ and ${\bf u}_k$, but note that the standard LQGR
formulation further restricts the form of the disturbance
distribution. For the estimation strategy to have a recursive solution,
we established in section \ref{DisStoCon} that each vector ${\bf v}_k$
can depend at most on ${\bf x}_k, {\bf x}_{k-1}, {\bf u}_{k-1}$ and
${\bf w}_{k-1}$, so the standard LQGR formulation further restricts the
noise distribution as well.  We will see below that, without most of
these additional restrictions, the nice analytical properties that arise
primarily because of the linearity of the models and Gaussianity of
the distributions are no longer present.

\subsubsection{Estimation Strategy (Kalman Filter)}
The standard LQGR formulation meets all of the criteria necessary to
apply the Kalman filtering algorithm \cite{Gel74:AppOE}. The main feature exploited by the
Kalman filter is that, because of the linearity of the system and
measurement models as well as the Gaussianity and independence of all
involved random variables, the probabilistic state $P_{x_k\mid I_k}$ at
each stage $k$ is guaranteed to remain Gaussian. Therefore, it is
sufficient to establish a recursive algorithm that yields the
second-order statistics
\begin{eqnarray*}
\begin{array}{rcl}
{\bf m}_{k\mid k} & = & E\left[x_k\mid I_k\right] \\ 
%{\bf \Sigma}_{k\mid k} & = & E\left[ \left.
%\left( {\bf x}_k - {\bf m}_{k\mid k}\right)'
%\left( {\bf x}_k - {\bf m}_{k\mid k}\right) \right| I_k \right]
{\bf \Sigma}_{k\mid k} & = & E\left[ \left.
\left( {\bf x}_k - {\bf m}_{k\mid k}\right)
\left( {\bf x}_k - {\bf m}_{k\mid k}\right)' \right| I_k \right]
\end{array}
%\label{eq:LQGRDefStats}
\end{eqnarray*}
for each stage $k$. Because the mean vector ${\bf m}_{k\mid k}$ minimizes expected-sum-square-error 
$$
E\left[ \left( x_k - \hat{x}_k\right)'\left( x_k - \hat{x}_k \right) \mid I_k\right]
$$ 
over estimates $\hat{x}_k$, it is often referred to as the {\it optimal state estimate}.
%The covariance matrix is a measure of the spread
%of the distribution about the mean; hence, the smaller this spread the
%more accurate the estimate. In the limit when the true state is known,
%${\bf m}_{k\mid k} = {\bf x}_k$ and ${\bf \Sigma}_{k\mid k} = {\bf 0}$
%(the zero matrix), corresponding to the pathological density of a unit
%impulse at the mean. 
Because the covariance matrix ${\bf \Sigma}_{k\mid k}$ has as its trace the expected-sum-equare-error of the optimal state estimate, it is often referred to as the {\it optimal error covariance}.

The Kalman filter is initialized according to
\begin{eqnarray}
\begin{array}{rcl}
{\bf \Sigma}_{0\mid 0} & = & {\bf \Sigma}_{x_0} - {\bf \Sigma}_{x_0}{\bf
C}_0'\left( {\bf C}_0 {\bf \Sigma}_{x_0}{\bf C}_0' + 
{\bf \Sigma}_{v_0}\right)^{-1} {\bf C}_0 {\bf \Sigma}_{x_0} \\[0.1in] 
{\bf m}_{0\mid 0} & = & {\bf m}_{x_0} + {\bf \Sigma}_{0\mid 0}{\bf
  C}_0'{\bf \Sigma}_{v_0}^{-1}
\left({\bf z}_0 - {\bf C}_0{\bf m}_{x_0} \right)
\end{array}
\label{eq:LQGRest0}
\end{eqnarray}
and note that these two equations characterize $\nu_0$, or the
estimation policy at stage $k=0$. Thereafter, the estimation policy
$\nu_k$ is characterized by each subsequent iteration of the
Kalman filter.  The {\it prediction step} relies on the system
equation $f_{k-1}$ to project how the conditional mean is expected to
evolve upon selection of control ${\bf u}_{k-1}$, or
\begin{eqnarray}
\begin{array}{rcl}
{\bf \Sigma}_{k \mid {k-1}} & = & {\bf A}_{k-1} {\bf \Sigma}_{{k-1}\mid {k-1}} 
{\bf A}_{k-1}' + {\bf \Sigma}_{w_{k-1}} \\[0.1in]
{\bf m}_{k\mid {k-1}} & = & {\bf A}_{k-1} {\bf m}_{{k-1}\mid {k-1}} + 
{\bf B}_{k-1} {\bf u}_{k-1}
\end{array} \qquad .
\label{eq:LQGRestP}
\end{eqnarray}
Upon receiving measurement $z_k$, the {\it correction step} then
relies on the measurement equation $h_k$ and adjusts the
prediction by an amount proportional to the difference between
the actual measurement and a predicted measurement, or
\begin{eqnarray}
\begin{array}{rcl}
{\bf \Sigma}_{k \mid k} & = & {\bf \Sigma}_{k \mid {k-1}} - 
{\bf \Sigma}_{k \mid {k-1}}{\bf
C}_k'\left( {\bf C}_k {\bf \Sigma}_{k \mid {k-1}}{\bf C}_k'
+ {\bf \Sigma}_{v_k}\right)^{-1} {\bf
C}_k {\bf \Sigma}_{k\mid {k-1}} \\[0.1in] 
{\bf m}_{k\mid k} & = & {\bf m}_{k\mid {k-1}} + 
{\bf \Sigma}_{k\mid k}{\bf C}_k'{\bf \Sigma}_{v_k}^{-1}
\left({\bf z}_k - {\bf C}_k{\bf m}_{k\mid {k-1}} - {\bf D}_{k-1}
{\bf u}_{k-1} \right)
\end{array} \qquad .
\label{eq:LQGRestC}
\end{eqnarray}
The weight on the corrective term is directly proportional to the
matrix ${\bf \Sigma}_{k\mid k}$ and inversely proportional to
the matrix ${\bf \Sigma}_{v_k}$. In essence, the weighting
balances between the accuracy of the prediction versus the accuracy of
the specific measurement; that is, the corrective term becomes
negligible if the uncertainty in the measurement is much greater than
that of the prediction.

Inspection of (\ref{eq:LQGRestP}) and (\ref{eq:LQGRestC}) suggests that
the measurements and controls only appear in the equations that generate
the mean vectors ${\bf m}_{k\mid k}$. It is surprising that the
estimation quality, as characterized by the covariance matrices ${\bf
\Sigma}_{k\mid k}$, does not depend on any element of the information
vector $I_k$ and can, therefore, be precomputed off-line.  This property
is arguably unique to the standard LQGR formulation---it certainly does
not hold in general. Even though the measurement equation $h_k$ depends
on the preceding control ${\bf u}_{k-1}$, a notion of sensor management
is effectively absent because the \underline{quality} of the estimate
remains unaffected. Thus, for the standard LQGR formulation, including
the term ${\bf D}_{k-1}{\bf u}_{K-1}$ in the measurement equation turns
out to be irrelevant with respect to capturing a notion of sensor 
management.

The most obvious extension of the standard LQGR formulation is to let
the noise covariance ${\bf \Sigma}_{v_k}$ be some function of control
$u_{k-1}$. For example, motivated by the PO-MDP formulation in
section~\ref{POMDP}, assume that $C_k = \Re^m \times
\{1,2\ldots,s_k\}$ and thus an additional discrete control variable,
call it $y_k$, determines whether one of $s_k$ possible noise
distributions characterizes the subsequent measurement statistics:
\begin{eqnarray}
{\bf z_k} = h_k({\bf x}_k, ({\bf u}_{k-1},y_{k-1}), {\bf v}_k) = 
{\bf C}_k {\bf x}_k + {\bf v}_k(y_{k-1}) \quad , \qquad k > 0 \qquad .
\label{eq:lqEx1meas}
\end{eqnarray}
Here, ${\bf v}_k(y_{k-1})$ still represents a zero-mean Gaussian
random vector but now we assume the covariance matrix is
control-dependent (i.e., ${\bf \Sigma}_{v_k(y_{k-1})}$ is different
for each of the $s_{k-1}$ values that $y_{k-1}$ can take). We augment
the cost equation for $k < K$ accordingly,
\begin{eqnarray}
  g_k({\bf x}_k, ({\bf u}_k,y_k), {\bf w}_k) = {\bf x}_k'{\bf
    T}_k{\bf x}_k + {\bf u}_k'{\bf R}_k{\bf u}_k + c_k(y_k)
\label{eq:lqEx1cost}
\end{eqnarray}
where the function $c_k:\{1,2,\ldots,s_k\} \rightarrow \Re$ defines
the ``relative expense'' of selecting from the set of measurement
equations---from a practical point-of-view, we should expect that
$c_k(o_1) > c_k(o_2)$ when the covariance ${\bf \Sigma}_{v_k(o_1)}$
yields a more accurate measurement equation than ${\bf
  \Sigma}_{v_k(o_2)}$.

As another example, we can let the noise covariance be an explicit
function of the control vector itself, or ${\bf \Sigma}_{v_k}({\bf
  u}_{k-1})$.  Assuming the functional form of ${\bf
  \Sigma}_{v_k}({\bf u}_{k-1})$ is such that higher-energy control
signals improve measurement accuracy, then a well-posed trade-off
arises between estimation performance (higher energy control reducing
estimation error penalties) and actuator performance (lower energy
control reducing quadratic cost penalties). Subsection~\ref{LQGR_Ex}
characterizes the impact that either suggested LQGR extension has on
the associated actuation strategy---we will discover that the
computational ramifications of the latter example exceed those of the
former example.

\subsubsection{Actuation Strategy}
The standard LQGR model, which as highlighted above does not represent
the opportunity for sensor management, admits a closed-form analytical
solution for the optimal actuation strategy. The optimal actuation
policy $\mu_k^*$ at each stage $k$ is a function of a special matrix
${\bf L}_k$, which can be precomputed off-line, and the mean vector output
of the recursive estimator ${\bf m}_{k \mid k}$. More specifically, the
optimal control at each stage is the result of the matrix multiplication
of ${\bf L}_k$ and ${\bf m}_{k\mid k}$, or
\begin{eqnarray}
\mu_k^*(P_{x_k \mid I_k}) = -{\bf L}_k {\bf m}_{k \mid k} \quad ,
\label{eq:LQGRAct}
\end{eqnarray}
and thus ${\bf L}_k$ is often called the {\it gain matrix}.  The matrix
${\bf L}_k$ is typically expressed in terms of another matrix ${\bf
K}_k$ that is the output of a backward recursion in $k$, each iteration
involving the matrices ${\bf A}_k$, ${\bf B}_k$, ${\bf T}_k$ and ${\bf
R}_k$ appearing in the system model and cost function. The recursion is
initialized by setting ${\bf K}_K = {\bf T}_K$ and then performing the
following recursive calculations for $k = K-1, K-2, \ldots, 0$:
\begin{eqnarray}
\begin{array}{rcl}
{\bf L}_k & = & \left( {\bf R}_k + {\bf B}_k'{\bf K}_{k+1}{\bf
B}_k \right)^{-1} {\bf B}_k'{\bf K}_{k+1}{\bf A}_k \\[0.1in]
{\bf P}_{k} & = & {\bf A}_k'{\bf K}_{k+1} {\bf B}_k {\bf L}_k \\[0.1in]
{\bf K}_k & = & {\bf A}_k'{\bf K}_{k+1}{\bf A}_k - 
{\bf P}_k + {\bf T}_k
\end{array} \qquad .
\label{eq:LQGRActRec}
\end{eqnarray}

The actuation strategy expressed by (\ref{eq:LQGRAct}) and
(\ref{eq:LQGRActRec}) depends on several properties of the standard
LQGR formulation that ensure each cost-to-go function preserves a
quadratic (convex) form in ${\bf u}_k$ and, in turn, leads to the
linearity of the actuation policy. Firstly, each Gaussian disturbance
${\bf w}_k$ has a covariance matrix that does not depend on control
vector ${\bf u}_k$ nor state vector ${\bf x}_k$, except possibly in a
linear or quadratic form. Secondly, the estimation error ${\bf
  \Sigma}_{k\mid k}$ cannot depend on the preceding control vector
${\bf u}_{k-1}$, again except possibly in a linear or quadratic form.
Note that the second statement is at direct odds with any notion of
sensor management! If the error covariance is a linear function of the
real-valued control vector ${\bf u}_{k-1}$, then we open the
possibility to a negative semi-definite covariance matrix, which is
senseless. This particular issue disappears if the error variance is
assumed to be a quadratic function of the control vector ${\bf
  u}_{k-1}$.  However, inspection of the Kalman filtering equations
suggests it is unlikely, assuming linear system/measurement models,
that one can find distributions leading to an estimator whose error
covariances at each stage $k$ will in fact preserve the desired
quadratic dependency on the control vector ${\bf u}_{k-1}$.

\subsection{Illustrative Examples \label{LQGR_Ex}}
Motivated by the previous subsection, we consider two extensions to
the standard LQGR formulation that each capture a notion of sensor
management.  The first extension augments the $m$-dimensional control
space with a finite set, each member of the set mapping to the same
linear-Gaussian measurement equation but with a different noise
covariance. We show that the control strategy still retains many of
its attractive features, with the most significant impact appearing in
the off-line computation for the estimation strategy; in particular,
the actuation strategy is unchanged and the estimation strategy is
still defined by a Kalman filter, but precomputation of the error
covariances now also requires the solution to a $K$-stage
\underline{non-stochastic} dynamic program.  The second extension
assumes the noise covariance is an explicit function of the
real-valued control vector ${\bf u }_{k-1}$.  Here, the error
covariances of the Kalman filter can no longer be precomputed
off-line.  Moreover, the cost-to-go function $J_{K-2}$ becomes a
non-convex function of the control vector ${\bf u}_{K-2}$ with
possibly multiple local minima, leading to a rather unattractive
optimization to solve for the actuation policy $\mu_{K-2}^*$ and, in
turn, limits the analytic tractability of any further iterations of
the main algorithm.  However, considering just a two-stage LQGR
problem with scalar state, control, disturbance, measurement and noise
spaces, we are able to computationally explore some interesting
properties related to such a sensor management scheme.

\subsubsection{Measurement Scheduling Using a Finite Set of Sensors}
We consider the standard LQGR formulation with the extension defined
by (\ref{eq:lqEx1meas}) and (\ref{eq:lqEx1cost}), in which the
opportunity for sensor management at each stage $k$ is modeled by an
additional discrete control variable $y_k \in \{1,2,\ldots,s_k \}$
whose value maps to one of a finite set of noise covariances ${\bf
  \Sigma}_{v_{k+1}(y_k)}$ in the equation characterizing the next
measurement ${\bf z}_{k+1}$. Starting with the main dynamic
programming algorithm, we will first conclude that control vectors
${\bf u}_k$ continue to be generated according to (\ref{eq:LQGRAct})
and (\ref{eq:LQGRActRec}). We also establish that, while the
estimation strategy also retains the same form as described by
(\ref{eq:LQGRest0})-(\ref{eq:LQGRestC}), the realized mean vectors
${\bf m}_{k|k}$ and error covariances ${\bf \Sigma}_{k|k}$ depend on
the previously selected controls $y_{k-1}, y_{k-2},\ldots,y_0$.
Perhaps surprisingly, we will also discover that the optimal {\it
  measurement schedule}, or the sequence of controls $y_0^*, y_1^*,
\ldots, y_{K-1}^*$, can be determined prior to on-line operation by
solving a particular deterministic dynamic program.  In summary, the
LQGR problem with measurement scheduling demands more off-line
computation than the standard LQGR problem, but on-line computation
turns out to be identical to the standard LQGR case.

The control at stage $k$ is the pair $({\bf u}_k,y_k) \in \Re^n \times
\{1,2,\ldots,s_k\} = C_k$. In the initialization stage of the
main algorithm, we have
\begin{eqnarray*}
J_{K-1}(I_{K-1}) & = & \min_{({\bf u}_{K-1},y_{K-1})} \left\{ E\left[ {\bf
      x}_{K-1}^\prime {\bf T}_{K-1} {\bf x}_{K-1} + {\bf u}_{K-1}^\prime 
      {\bf R}_{K-1} {\bf u}_{K-1} + c_{K-1}(y_{K-1}) + \right. \right. \\ 
& & \qquad \qquad \qquad \left( {\bf A}_{K-1}{\bf x}_{K-1} + 
{\bf B}_{K-1}{\bf u}_{K-1} + {\bf w}_{K-1}\right)^\prime {\bf T}_K \\ 
& & \qquad \qquad \qquad \quad \left. \left. 
\left( {\bf A}_{K-1}{\bf x}_{K-1} + {\bf B}_{K-1}{\bf u}_{K-1} + 
{\bf w}_{K-1}\right) \mid I_{K-1},{\bf u}_{K-1},y_{K-1}\right] \right\} 
\\[0.1in]
& = & E\left[ {\bf x}_{K-1}^\prime \left( {\bf A}_{K-1}^\prime {\bf T}_K
{\bf A}_{K-1} + {\bf T}_{K-1} \right) {\bf x}_{K-1} \mid I_{K-1} \right] + \\
& & E\left[ {\bf w}_{K-1}^\prime {\bf T}_K {\bf w}_{K-1} \mid I_{K-1} 
\right] + \\ & & \min_{{\bf u}_{K-1}} \left\{ {\bf u}_{K-1}^\prime
\left( {\bf B}_{K-1}^\prime {\bf T}_K {\bf B}_{K-1} + {\bf R}_{K-1}
\right){\bf u}_{K-1} + \right. \\
& & \qquad \quad \left. 2E\left[ {\bf x}_{K-1}^\prime\mid I_{K-1}\right] 
{\bf A}_{K-1}^\prime {\bf T}_K {\bf B}_{K-1} {\bf u}_{K-1} \right\} + 
\min_{y_{K-1}} c_{K-1}(y_{K-1})
\end{eqnarray*}
where we have used the fact that disturbance vector ${\bf w}_{K-1}$
is zero-mean and independent of state vector ${\bf x}_{K-1}$. The
expression for cost-to-go function $J_{K-1}$ is the same as in the
conventional LQGR problem except, of course, for the last term.  The
optimal actuation policy $\mu_{K-1}^*$, then, generates controls
according to
$$
{\bf u}_{K-1}^* = -{\bf L}_{K-1} E\left[ {\bf x}_{K-1} \mid
  I_{K-1}\right] \qquad \mbox{and} \qquad y_{K-1}^* = \arg\min_{y_{K-1}}
c_{K-1}(y_{K-1})
$$
where ${\bf L}_{K-1}$ is the same gain matrix used in the standard
LQGR solution. Because a measurement is not used in terminal stage
$K$, it is no surprise that the scheduling decision in stage $K-1$ is
based solely on the measurement cost function $c_{K-1}$.  In stages $k
< K-1$, however, we should expect that the measurement accuracy as
expressed by noise covariance ${\bf \Sigma}_{v_{k+1}(y_{k+1})}$ will
somehow enter into consideration.

Though we will omit the steps here, it is straightforward (albeit
tedious, too) to apply the next stage of the dynamic programming
algorithm. Doing so shows that the minimizing scheduling variable
$y_{K-2}^*$ satisfies
\begin{eqnarray*}
& & \min_{y_{K-2}} \left\{ E\left[ 
E\left[ \left( {\bf x}_{K-1} - E\left[
          {\bf x}_{K-1} \mid I_{K-1}\right]\right)^\prime {\bf
        P}_{K-1} \left( {\bf x}_{K-1} - E\left[ {\bf x}_{K-1} \mid
          I_{K-1}\right]\right) \mid I_{K-1}\right] \right. \right. \\ 
 & & \qquad \qquad \qquad \qquad \qquad \qquad 
\qquad \qquad \qquad \qquad \qquad
\left. \left. \mid I_{K-2},{\bf u}_{K-2},y_{K-2} \right] + 
c_{K-2}(y_{K-2})\right\}
\end{eqnarray*}
where we recognize the inner conditional expectation as the (weighted)
error covariance ${\bf \Sigma}_{K-1\mid K-1}$ of the state estimate
$E[{\bf x}_{K-1}\mid I_{K-1}]$ and the outer conditional expectation
is with respect to ${\bf z}_{K-1}$. No matter how we choose $y_{K-2}$,
linearity of the models guarantees that vectors ${\bf x}_{K-1}$ and
${\bf z}_{K-1}$ remain Gaussian and thus the Kalman filter equations
still characterize the estimation strategy. As before, the Kalman
filter's error covariance does not depend on control vector ${\bf
  u}_{K-2}$ nor on measurement vector ${\bf z}_{K-1}$, reducing the
minimization to
\begin{equation}
\begin{array}{l}
\displaystyle 
\min_{y_{K-2}} \left\{ E\left[ \left( {\bf x}_{K-1} - E\left[ {\bf
          x}_{K-1} \mid I_{K-1}\right]\right)^\prime {\bf P}_{K-1}
    \left( {\bf x}_{K-1} - E\left[ {\bf x}_{K-1} \mid
        I_{K-1}\right]\right) \mid I_{K-2},y_{K-2} \right] + \right. \\
\left. \qquad \qquad \qquad \qquad \qquad \qquad \qquad \qquad \qquad
\qquad \qquad \qquad \qquad \qquad \qquad \qquad c_{K-2}(y_{K-2})\right\}
\end{array}
\label{eq:LQGRex1estDP1}
\end{equation}
The other terms of $J_{K-2}$ in which the control vector ${\bf
  u}_{K-1}$ appears are identical to the terms in the standard LQGR
case and, therefore, the minimizing control vector continues to be
$$
{\bf u}_{K-2}^* = -{\bf L}_{K-2} E\left[ {\bf x}_{K-2} 
\mid I_{K-2}\right] \qquad .
$$

Recall that in the standard LQGR solution, the Kalman filter's error
covariances could be readily precomputed via recursions
(\ref{eq:LQGRest0})-(\ref{eq:LQGRestC}), each iteration completely
determined by the specified models.  Now, however, the correction step
to generate each error covariance ${\bf \Sigma}_{k|k}$ for $k > 0$
depends on noise covariance ${\bf \Sigma}_{v_k(y_{k-1})}$ and thereby
also on scheduling decision $y_{k-1}$---we notate this dependence with
a superscript $y$ on each error covariance.  Given the previous error
covariance ${\bf \Sigma}^y_{k|k}$ and the previous measurement
decision $y_{k}$, we define function $\tilde{f}_k$ as shorthand for
the error covariance recursion i.e., ${\bf \Sigma}^y_{k+1|k+1} =
\tilde{f}_k\left( {\bf \Sigma}^y_{k|k}, y_k \right)$ denotes
\begin{equation}
\begin{array}{rcl}
{\bf \Sigma}_{k+1 \mid k} & = & {\bf A}_k 
{\bf \Sigma}^y_{k\mid k} 
{\bf A}_k' + {\bf \Sigma}_{w_k} \quad \mbox{and} \\[0.1in]
{\bf \Sigma}^y_{k+1 \mid k+1} & = & {\bf \Sigma}_{k+1 \mid k} - 
{\bf \Sigma}_{k+1 \mid k}{\bf
C}_{k+1}'\left( {\bf C}_{k+1} {\bf \Sigma}_{k+1 \mid k}{\bf C}_{k+1}'
+ {\bf \Sigma}_{v_{k+1}(y_k)}\right)^{-1} {\bf
C}_{k+1} {\bf \Sigma}_{k+1\mid k}
\end{array}
\label{eq:LQGRex1meaSys}
\end{equation}
for $k = 0, 1, \ldots, K-2$. Also employing a well-known identity for
a weighted covariance matrix, we can express (\ref{eq:LQGRex1estDP1})
as
\begin{eqnarray*}
\min_{y_{K-2}} \left\{ c_{K-2}(y_{K-2}) + \mbox{trace} 
\left({\bf P}_{K-1}^{\frac{1}{2}}
      \tilde{f}_{K-2} \left( {\bf \Sigma}_{K-2|K-2}^y,y_{K-2} \right)
{\bf P}_{K-1}^{\frac{1}{2}}\right) \right\} \quad .
\end{eqnarray*}
This minimization is clearly dependent on error covariance ${\bf
  \Sigma}^y_{K-2|K-2}$, resembling a cost-to-go function in which the
previous error covariance ${\bf \Sigma}^y_{K-2|K-2}$ assumes the role
of ``state'' and the measurement decision $y_{K-2}$ assumes the role
of ``control.'' Accordingly, we define the optimal {\it measurement
  policy} $\theta^*_{K-2}$ at stage $K-2$ by the function
\begin{equation}
\theta_{K-2}^*\left({\bf \Sigma}^y_{K-2|K-2}\right) = 
\arg\min_{y_{K-2}} \left\{ c_{K-2}(y_{K-2}) + 
  \mbox{trace} \left({\bf P}_{K-1}^{\frac{1}{2}}\tilde{f}_{K-2}\left({\bf
      \Sigma}_{K-2|K-2}^y,y_{K-2}\right){\bf P}_{K-1}^{\frac{1}{2}}\right) 
\right\} \quad .
\label{eq:LQGRex1actM}
\end{equation}
Note that such a measurement policy is also well-defined in stage
$K-1$ i.e.,
$$
\theta^*_{K-1}({\bf \Sigma}_{K-1|K-1}^y) = \arg\min_{y_{K-1}} \left\{
c_{K-1}(y_{K-1}) \right\} \qquad ,
$$
even though this generality is unnecessary due to the absence of a
next measurement and, in turn, the absence of actual dependence on the
final error covariance ${\bf \Sigma}_{K-1|K-1}^y$.

Subsequent iterations of the main dynamic programming algorithm
develop in exactly the same way. The optimal control vector ${\bf
  u}^*_k$ gets generated according to (\ref{eq:LQGRAct}) and the
optimal measurement decision $y^*_{k}$ according to a measurement
policy $\theta^*_k$ that is analogous to that described by
(\ref{eq:LQGRex1actM}) for stage $K-2$.  Note the decoupling in the
off-line computation of these two policies.  Moreover, neither policy
is dependent on the realization of measurement vector ${\bf z}_k$;
during on-line operation, of course, the measurement vector affects
the state estimate ${\bf m}_{k|k}$ and, in turn, impacts the
particular control vector ${\bf u}_k^*$ generated by that policy. The
measurement decision, on the other hand, depends only on the error
covariances, which in the Kalman filter do \underline{not} depend on
the realized measurements. This means that the opportunity for
measurement scheduling only impacts off-line computation; in
particular, once the following (deterministic) dynamic programming
algorithm is solved to obtain the optimal measurement schedule $y_0^*,
y_1^*, \ldots, y_{K-1}^*$, the
problem is effectively reduced to a standard LQGR formulation. \\
\begin{minipage}[h]{\textwidth}
\rule{\textwidth}{0.01in}
\begin{footnotesize}
  {\bf Measurement Scheduling Algorithm:} Let matrices ${\bf P}_k$ be
  computed according to (\ref{eq:LQGRActRec}) and define functions
  $\tilde{f}_k$ by (\ref{eq:LQGRex1meaSys}). For every initial error
  covariance ${\bf \Sigma}_{0|0}$, the optimal total measurement cost
  is equal to $H_0\left({\bf \Sigma}_{0|0}\right)$, where the function
  $H_0$ is given by the last step of the following algorithm:

\begin{description}
\item[\it Initialization:] Given ${\bf \Sigma}^y_{K-1|K-1}$, define
\begin{eqnarray*}
H_{K-1}\left({\bf \Sigma}^y_{K-1|K-1}\right) & = & 
\mbox{trace}\left( {\bf P}_{K-1}^\frac{1}{2} {\bf \Sigma}^y_{K-1|K-1}
{\bf P}_{K-1}^\frac{1}{2}\right) +
\min_{y_{K-1} \in \{1,\ldots,s_{K-1}\}} \left\{ c_{K-1}(y_{K-1}) \right\} 
\qquad .
\end{eqnarray*}

\item[\it Recursion:] For stages $k = K-2, K-3, \ldots, 0$, define
\begin{eqnarray*}
H_k\left({\bf \Sigma}^y_{k|k}\right) & = & 
\mbox{trace}\left( {\bf P}_k^\frac{1}{2} {\bf \Sigma}^y_{k|k}
{\bf P}_k^\frac{1}{2}\right) +
\min_{y_k \in \{1,\ldots,s_k\}} \left\{ c_k(y_k) + H_{k+1}\left( 
\tilde{f}({\bf \Sigma}^y_{k|k},y_k)\right)\right\} \qquad .
\end{eqnarray*}
\end{description}
At each iteration $k$, the minimization that yields $H_k$ for each
${\bf \Sigma}^y_{k|k}$ simultaneously defines the associated optimal
measurement policy $\theta^*_k$; that is, if $y_k^* = \theta^*_k({\bf
  \Sigma}^y_{k|k})$ minimizes $H_k( {\bf \Sigma}^y_{k|k})$ for each
possible ${\bf \Sigma}^y_{k|k}$, then the resulting measurement
strategy $(\theta^*_0,\theta^*_1, \ldots, \theta^*_{K-1})$ is optimal.
Because the initial error covariance ${\bf \Sigma}_{0|0}$ is
determined exclusively by the initial state distribution and initial
measurement model via (\ref{eq:LQGRest0}), forward application of this
optimal measurement strategy combined with the sequence of known
functions $\tilde{f}_0, \tilde{f}_1, \ldots, \tilde{f}_{K-1}$ realizes
the optimal measurement schedule $y^*_0,y^*_1,\ldots, y^*_{K-1}$. 
\\[-0.1in]
\end{footnotesize}
\rule{\textwidth}{0.01in}
\end{minipage}

\subsubsection{Direct Control of Measurement Accuracy}
The preceding example modeled the opportunity for sensor management by
augmenting the control space at each stage to include the selection of
one out of a finite set of sensors to generate the next measurement.
We concluded that this scheme preserves all of the desirable
computational properties of LQGR problems. A distinct alternative is
to let the control vector ${\bf u}_k$ directly influence the quality
of the next measurement, corresponding in some sense to selection from
an infinite set of sensors at each stage. We will assume that more
costly controls, as measured by the quadratic cost function, lead to
more accurate measurements. In this section, we consider the standard
LQGR problem except where the noise covariance matrix ${\bf
  \Sigma}_{v_{k+1}}$ is a function of the preceding control vector
${\bf u}_k$ such that, for each $k = 0, 1,\ldots, K-1$,
$$
{\bf u}_k^\prime {\bf u}_k < \bar{\bf u}_k^\prime
\bar{\bf u}_k \quad \Longleftrightarrow \quad \mbox{trace}\left({\bf
  \Sigma}_{v_{k+1}}({\bf u}_k)\right) > \mbox{trace}\left({\bf
  \Sigma}_{v_{k+1}}(\bar{\bf u}_k)\right) > 0 \qquad .
$$
In words, the above condition says ``under imperfect state
information, lower-energy control implies higher noise power in the
subsequent measurement equation.''  Not surprisingly, we will discover
that such a sensor management scheme has more severe computational
implications than the scheme in the preceding example.

\paragraph{The Model:} To most clearly illustrate the impact of the proposed
sensor management scheme, we limit the analysis of this section to the
simplest instance of an LQGR problem: a stationary system, cost, and
measurement equation with scalar state, control and measurement spaces
over just two decision stages. More specifically, we assume $K = 2$,
structural parameters
$$
n = m = s = 1 \quad \Rightarrow \quad S_k=C_k=Z_k=\Re,
$$
linear system equations
$$
x_{k+1} = a x_k + b u_k + w_k \: , \quad {\bf \Sigma}_{w_k} =
\sigma_w^2 > 0
$$
with initial state distribution given by ${\bf m}_{x_0} = m_x$ and
${\bf \Sigma}_{x_0} = \sigma_x^2 > 0$, quadratic cost equations
$$
tx_k^2 + ru_k^2 \:, \quad t \geq 0 \mbox{ and } r > 0
$$
with terminal cost $tx_2^2$, and linear measurement equations
\begin{eqnarray*}
  z_0 & = & cx_0 + v_0 \: , \quad {\bf \Sigma}_{v_0} = \sigma_v^2 > 0 \\
  z_1 & = & cx_1 + du_0 + v_1 \: , \quad {\bf \Sigma}_{v_1}({\bf
      u}_0) = \frac{\sigma_v^2}{1 + \gamma u_0^2}
\mbox{ for some fixed } \gamma > 0 \quad . 
\end{eqnarray*}
We emphasize that the value of $\gamma$ parameterizes the degree to
which the opportunity for sensor management is present. As $\gamma$
approaches zero, the above problem approaches its standard LQGR
counterpart whereas, for a fixed applied control $u_0$, increasing
parameter $\gamma$ induces a more accurate measurement $z_1$. Indeed, we
should expect the degree of {\it sensor influence} captured by parameter
$\gamma$ to manifest itself, in some manner, in the optimal solution for
this two-stage problem.

\paragraph{The Estimation Strategy:} Recall that the control $u_0$ is
part of the information state $I_1$. Thus, even with the
control-dependent distribution for measurement noise $v_1$, the
linearity of the system and measurement models preserves Gaussianity of
the probabilistic state estimate, implying the Kalman filter recursions
continue to apply. For the scalar two-stage model in question,
(\ref{eq:LQGRest0})-(\ref{eq:LQGRestC}) specialize to:
\begin{eqnarray*}
\mbox{\it Initialization:} & \quad & 
\begin{array}{rcl}
{\bf \Sigma}_{0|0} & = & \displaystyle 
\frac{\sigma^2_x\sigma_v^2}{c^2\sigma_x^2 + \sigma_v^2} \quad \equiv 
\quad \sigma^2_{0|0} \\[0.2in]
{\bf m}_{0|0} & = & \displaystyle 
m_x + c\left( \frac{\sigma^2_{0|0}}{\sigma^2_v}\right)
\left( z_0 - cm_x\right) \quad \equiv \quad m_{0|0} 
\end{array} \\[0.2in]
\mbox{\it Prediction:} & &
\begin{array}{rcl}
{\bf \Sigma}_{1|0} & = & a^2\sigma_{0|0}^2 + \sigma_w^2 \quad \equiv 
\quad \sigma^2_{1|0} \\[0.1in]
{\bf m}_{1|0} & = & am_{0|0} + bu_0 \quad \equiv \quad m_{1|0}(u_0) 
\end{array} \\[0.2in]
\mbox{\it Correction:} & &
\begin{array}{rcl}
{\bf \Sigma}_{1|1} & = & \displaystyle 
\frac{\sigma^2_{1|0}\sigma^2_v}{c^2\sigma^2_{1|0}(1+\gamma u_0^2)+
\sigma_v^2} \quad \equiv \quad \sigma_{1|1}^2(u_0) \\[0.3in]
{\bf m}_{1|1} & = & \displaystyle m_{1|0}(u_0) + 
c\left(\frac{\sigma_{1|1}^2(u_0)(1+\gamma u_0^2)}{\sigma^2_v}\right)
(z_1 - cm_{1|0}(u_0) - du_0) \\
& \equiv & m_{1|1}(u_0) 
\end{array}
\end{eqnarray*}
Our notation in the predicted mean and both parts of the correction
step reminds us that these quantities depend on control $u_0$. In the
standard LQGR case, only a linear dependence on control $u_0$ would
appear; in our case, the dependence is nonlinear! It is also worth
noting that, all other things equal, a larger value of $\gamma$ yields
a smaller error variance $\sigma^2_{1|1}(u_0)$ and, in turn, places a
higher weight on realized measurement $z_1$ in the computation of
state estimate $m_{1|1}(u_0)$.

\paragraph{The Actuation Strategy:} As was shown in the measurement 
scheduling example, the initialization stage of the main dynamic
programming algorithm is unaffected by the opportunity for sensor
management. The matrix computations of (\ref{eq:LQGRActRec}) become
$$
\begin{array}{rcl}
{\bf L}_1 & = & \displaystyle \frac{bta}{r+b^2t} \quad 
\equiv \quad \ell_1 \\[0.2in]
{\bf P}_1 & = & \displaystyle bta\ell_1 = \frac{(bta)^2}{r+b^2t} \quad 
\equiv \quad \rho_1 \\[0.2in]
{\bf K}_1 & = & \displaystyle a^2t - \rho_1 + t = 
\frac{t(a^2r + b^2t +r)}{r+b^2t} \quad \equiv \quad \kappa_1
\end{array} \qquad , 
$$
leading to optimal actuation policy via (\ref{eq:LQGRAct}) of
$$
\mu^*_1(P_{x_1|I_1}) = -\ell_1 m_{1|1}(u_0)
$$
and associated optimal cost-to-go function, after some algebra, of
\begin{equation}
J_1(P_{x_1|I_1}) = (\kappa_1+\rho_1)\sigma^2_{1|1}(u_0) +
\kappa_1[m_{1|1}(u_0)]^2 + t \sigma^2_w \qquad .
\label{eq:LQGRex2J1}
\end{equation}
The next step of the dynamic programming algorithm is to compute
cost-to-go function $J_0$, which will involve a minimization of $J_1$
(as well as some other additive terms) over $u_0$. It is easy to see
that $J_1$ can depend on control $u_0$ in a non-convex way, so we should
\underline{not} expect the optimal policy at stage $k=0$ to also match
the standard LQGR case.

We now apply the dynamic programming algorithm for stage $k=0$,
starting with the computation of the function $Q_0$:
\begin{eqnarray*}
Q_0\left(P_{x_0|I_0},u_0\right) & = & E\left[ E\left[ tx_0^2 + ru_0^2 + 
J_1\left(\nu_1(P_{x_0|I_0},u_0,z_1)\right)|x_0,u_0\right] |I_0 \right] \\
& = & tE[x_0^2|I_0] + ru_0^2 + E\left[ J_1\left(\nu_1(P_{x_0|I_0},u_0,z_1) 
\right)|I_0,u_0\right] \qquad .
\end{eqnarray*}
Given only $I_0$ and $u_0$, cost-to-go $J_1$ will be a function of
$z_1$. We substitute into (\ref{eq:LQGRex2J1}) the estimation equations 
that depend explicitly on $u_0$ and $z_1$ and obtain
\begin{eqnarray*}
J_1\left(\nu_1(P_{x_0|I_0},u_0,z_1)\right) & = &
(\kappa_1 + \rho_1)\left( \frac{\sigma^2_{1|0}\sigma_v^2}
{\alpha} \right) + \kappa_1 \left[ m_{1|0}(u_0) + c\left( 
\frac{\sigma^2_{1|0}(1+\gamma u_0^2)}{\alpha}\right)\beta \right]^2 + 
t\sigma^2_w \\[0.1in]
& = & \frac{1}{\alpha^2}
\left[ \left(t\sigma^2_w + \kappa_1[m_{1|0}(u_0)]^2\right)\alpha^2 + 
(\kappa_1 + \rho_1)\sigma^2_{1|0}\sigma^2_v \alpha + \right. \\ 
& & \qquad \qquad \left. 2c\sigma^2_{1|0}(1+\gamma u_0^2)m_{1|0}(u_0)\alpha 
\beta + c^2 \sigma^4_{1|0} (1+\gamma u_0^2)^2 \beta^2 \right] 
\end{eqnarray*}
where we have defined intermediate variables 
$$
\alpha =
c^2\sigma^2_{1|0}(1+\gamma u_0^2) + \sigma_v^2 \qquad \mbox{and} \qquad  
\beta = z_1-cm_{1|0}(u_0)-du_0 \qquad .
$$
Note, firstly, that conditioned on $I_0$ and $u_0$, variable $\alpha$
is guaranteed to be a positive constant and, secondly, that
measurement $z_1$ is absent from $J_1$ except within the variable
$\beta$. The latter implies that computing the conditional expectation
of $J_1$ boils down to computing the conditional expectation of
$\beta$ and $\beta^2$:
\begin{eqnarray*}
E\left[ \beta | I_0,u_0\right] & = & E\left[ z_1 - cm_{1|0}(u_0) - du_0 
|I_0,u_0 \right] \\ 
& = & E\left[ cx_1+du_0 + v_1|I_0,u_0 \right] - cm_{1|0}(u_0) - du_0 \\ 
& = & cm_{1|0}(u_0) + du_0 - cm_{1|0}(u_0) - du_0 = 0 \\[0.1in]
E\left[ \beta^2 | I_0,u_0\right] & = & \mbox{var}(\beta|I_0,u_0) = 
\mbox{var}(z_1 - cm_{1|0}(u_0) - du_0|I_0,u_0) \\ & = & 
\mbox{var}(cx_1+du_0+v_1|I_0,u_0) = c^2\mbox{var}(x_1|I_0,u_0) + 
\mbox{var}(v_1|I_0,u_0) \\ & = &
c^2\sigma^2_{1|0} + \frac{\sigma_v^2}{1+\gamma u_0^2} = 
\frac{c^2\sigma^2_{1|0}(1+\gamma u_0^2) + \sigma_v^2}{1+\gamma u_0^2} = 
\frac{\alpha}{1+\gamma u_0^2}
\end{eqnarray*}
It follows that 
\begin{eqnarray*}
E\left[ J_1\left(\nu_1(P_{x_0|I_0},u_0,z_1) \right)|I_0,u_0 \right] & = & 
\frac{1}{\alpha}\left[ \left(t\sigma^2_w + \kappa_1[m_{1|0}(u_0)]^2
\right)\alpha + \right. \\ & & \qquad \qquad \qquad \qquad \left. 
(\kappa_1 + \rho_1)\sigma^2_{1|0}\sigma^2_v + c^2 \sigma^4_{1|0} 
(1+\gamma u_0^2) \right] \quad ,
\end{eqnarray*}
which combined with
$$
E[x_0^2|I_0] = \mbox{var}(x_0|I_0) + \left(E[x_0|I_0]\right)^2 = 
\sigma^2_{0|0} + m_{0|0}^2 \quad \mbox{and} \quad [m_{1|0}(u_0)]^2 = 
a^2m^2_{0|0} + 2abm_{0|0}u_0 + b^2u_0^2
$$
yields
\begin{eqnarray*}
Q_0(P_{x_0|I_0},u_0) & = & t(\sigma^2_{0|0} + m_{0|0}^2) + ru_0^2 + 
\frac{1}{\alpha}\left[ \left(t\sigma^2_w + \kappa_1(a^2m^2_{0|0} + 
2abm_{0|0}u_0 + b^2u_0^2) \right)\alpha + \right. \\[0.1in] 
& & \qquad \qquad \qquad \qquad \qquad \qquad \left. 
(\kappa_1 + \rho_1)\sigma^2_{1|0}\sigma^2_v + c^2 \sigma^4_{1|0} 
(1+\gamma u_0^2) \right] \quad ,
\end{eqnarray*}
Finally, substituting the expression for $\alpha$ and consolidating
terms over a common denominator, we can express the control-dependent
cost-to-go function $Q_0$ as a ratio of polynomials in $u_0$:
\begin{equation}
\begin{array}{rcl}
\qquad Q_0\left(P_{x_0|I_0},u_0\right) & = & \displaystyle 
\frac{\alpha_4u_0^4+\alpha_3u_0^3+\alpha_2u_0^2+\alpha_1u_0 +\alpha_0}
{\beta_2 u_0^2 + \beta_0} \\[0.2in]
\multicolumn{3}{l}{\mbox{where}} \\[0.1in]
\beta_2 & = & c^2\sigma^2_{1|0}\gamma \\[0.1in]
\beta_0 & = & c^2\sigma^2_{1|0} + \sigma^2_v \\[0.1in]
\alpha_4 & = & (r+b^2\kappa_1)\beta_2 \\[0.1in]
\alpha_3 & = & 2ab\kappa_1m_{0|0}\beta_2 \\[0.1in]
\alpha_2 & = & (r+b^2\kappa_1)\beta_0 + \left(t(m^2_{0|0} + \sigma^2_{0|0} + 
\sigma^2_w)+\kappa_1a^2m^2_{0|0}\right)\beta_2 + 
c^2\sigma^4_{1|0}\gamma \\[0.1in]
\alpha_1 & = & 2ab\kappa_1m_{0|0} \beta_0 \\[0.1in]
\alpha_0 & = & \left(t(m^2_{0|0} + \sigma^2_{0|0} + \sigma^2_w) + 
\kappa_1a^2m^2_{0|0} \right)\beta_0 + (\kappa_1 + \rho_1) 
\sigma_{1|0}^2\sigma_v^2 + c^2 \sigma^4_{1|0}
\end{array} .
\label{eq:LQGRex2Q0}
\end{equation} 

As a sanity check, consider (\ref{eq:LQGRex2Q0}) as $\gamma$
approaches zero.  Coefficient $\beta_2$ approaches zero and, in turn,
coefficients $\alpha_4$ and $\alpha_3$ also approach zero and coefficient 
$\alpha_2$ approaches $(r+b^2\kappa_1)\beta_0$. Equation (\ref{eq:LQGRex2Q0}) 
reduces to the quadratic function of $u_0$, 
\begin{eqnarray*}
\lim_{\gamma \rightarrow 0} Q\left(P_{x_0|I_0},u_0\right) & = &
(r+b^2\kappa_1)u_0^2 + 2ab\kappa_1m_{0|0}u_0 + \alpha_0 \qquad ,
\end{eqnarray*}
agreeing with known behavior of the standard LQGR solution. Taking the
derivative with respect to $u_0$ and equating to zero results in
$$
\lim_{\gamma \rightarrow 0} u^*_0 =
-\left(\frac{ab\kappa_1}{r+b^2\kappa_1}\right)
m_{0|0} = -\ell_0 m_{0|0} \qquad ,
$$
also agreeing with the solution in (\ref{eq:LQGRAct}).  Note that
when $\gamma>0$, the higher-order terms in the numerator and
denominator polynomials of $Q_0$ are also nonzero, and in fact
increase as $\gamma$ increases. At the point that these higher-order
terms dominate the lower-order terms, the desirable convexity property
of the quadratic expression is no longer guaranteed.

\begin{figure}[t!]
\begin{center}
  {\bf Parameters:} $(m_x,\sigma_x,a,b,\sigma_w,c,d,\sigma_v) =
  (0,1,1,1,1,1,0,1)$, \quad $(t,r) = (1,1)$ \\[0.2in]
\resizebox{0.75\textwidth}{!}{\includegraphics{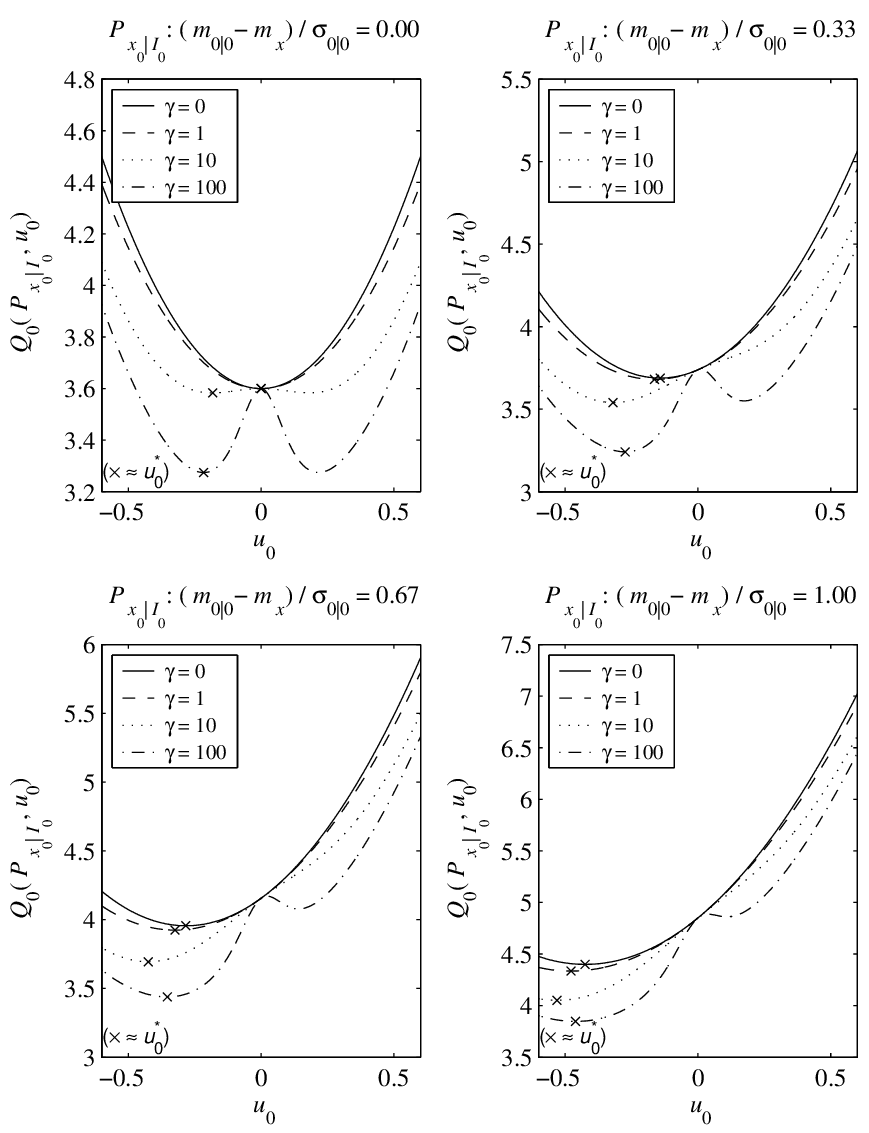}}
\caption{Control-Dependent Cost-To-Go Function; 
  Varying Degrees of Sensor Influence}
\label{fig:lqgrEx1}
\end{center}
\end{figure}

Figure~\ref{fig:lqgrEx1} visualizes the function $Q_0$, and its
dependence on model parameter $\gamma$, assuming all other model
parameters are fixed. Each plot corresponds to a different case of the
probabilistic state estimate; the first plot reflects a measurement
$z_0$ exactly at its predicted value $cm_x$, implying no influence in
the correction step; the second plot reflects a measurement $z_0$
greater than its predicted value such that the initialization step
results in an estimate $m_{0|0}$ that is one-third of a standard
deviation $\sigma_{0|0}$ from the initial state estimate $m_x$; the
third and fourth plots are similar, except with two-thirds and a whole
standard deviation, respectively. (Though not shown, measurements
$z_0$ below the predicted value produce mirror images, about the $u_0
= 0$ axis, of their above-the-predicted-value counterparts.) For any
fixed $u_0$, increasing $\gamma$ leads to decreasing (or at least
non-increasing) cost-to-go, reflecting the notion that a more accuate
measurement improves control performance, all other things equal. Also
note that, at $u_0 = 0$, the curves pass through the same point
regardless of the value of $\gamma$---this is in clear agreement with
our choice for the functional form ${\bf \Sigma}_{v_1}({\bf u}_0)$.
For the assumed model parameters, the non-convexity of $Q_0$ versus
$u_0$ is especially apparent for $\gamma = 10$ and $\gamma = 100$.  On
all curves, the symbol $\times$ indicates (approximately) the global
minimum; notice that $u_0^*$ is \underline{not} monotonic with
$\gamma$. The intuition is that, for a fixed $\gamma$, there is a
value of $u_0$ above which the marginal improvement in measurement
accuracy is negligible, at which point there is no longer any
incentive to incur the additional penalty measured by the quadratic
cost.

Having obtained the function $Q_0$, we now proceed to solve for the
optimal actuation policy $\mu_0$. Given any probabilistic state
$P_{x_0|I_0}$, the minimizing control $u_0^*$ necessarily satisfies
the basic derivative test:
$$
\frac{dQ_0\left(P_{x_0|I_0},u_0\right)}{du_0} = 0 \qquad \mbox{and} \qquad 
\frac{d^2Q_0\left(P_{x_0|I_0},u_0\right)}{du_0^2} > 0 \qquad .
$$
Of course, without the convexity assumption, it is possible that
these equations are satisfied by multiple values of $u_0$, in which case the
global minimum $u_0^*$ is such a $u_0$ for which evaluation of $Q_0$
is also smallest. Computing the first derivative,
\begin{eqnarray*}
\frac{dQ_0\left(P_{x_0|I_0},u_0\right)}{du_0} & = & \frac{\begin{array}{lcr}
\left( 4\alpha_4u_0^3 + 3\alpha_3u_0^2 + 2\alpha_2u_0 + \alpha_1\right)
\left(\beta_2u_0^2 + \beta_0\right) & - & \qquad \qquad \qquad \qquad \\
\multicolumn{3}{r}{\left( \alpha_4u_0^4+ \alpha_3u_0^3+\alpha_2u_0^2+
\alpha_1u_0 +\alpha_0 \right)2\beta_2u_0} \end{array}}
{\left(\beta_2u_0^2 + \beta_0\right)^2} \\[0.1in]
& = & \frac{\delta_5u_0^5 + \delta_4u_0^4 + \delta_3u_0^3 + \delta_2u_0^2 
+ \delta_1u_0 + \delta_0}{\left(\beta_2u_0^2+\beta_0\right)^2} \\
\mbox{where} \qquad \qquad & & \\
\delta_5 & = & 2\alpha_4\beta_2 \\
\delta_4 & = & \alpha_3\beta_2 \\
\delta_3 & = & 4\alpha_4\beta_0 \\
\delta_2 & = & \left( 3\alpha_3\beta_0 - \alpha_1\beta_2\right) \\
\delta_1 & = & 2\left( \alpha_2\beta_0 - \alpha_0\beta_2\right) \\
\delta_0 & = &\alpha_1\beta_0
\end{eqnarray*}
from which the second derivative follows similarly:
$$
\begin{array}{rcl} \displaystyle 
\frac{d^2Q_0\left(P_{x_0|I_0},u_0\right)}{du_0^2} & = & \displaystyle 
\frac{\epsilon_8u_0^8 + \epsilon_6u_0^6 + \epsilon_5u_0^5 + 
\epsilon_4u_0^4 + \epsilon_3u_0^3 + \epsilon_2u_0^2 + \epsilon_1u_0 + 
\epsilon_0}{\left(\beta_2u_0^2 + \beta_0\right)^4} \\[0.1in]
\mbox{where} \qquad \qquad \qquad \qquad & & \\
\epsilon_8 & = & \delta_5\beta_2^2 \\[0.1in]
\epsilon_6 & = & 6\delta_5\beta_2\beta_0 - \delta_3\beta_2^2 \\[0.1in]
\epsilon_5 & = & 4 \delta_4\beta_2\beta_0 - 2\delta_2\beta_2^2 \\[0.1in]
\epsilon_4 & = & 5\delta_5\beta_0^2 + 2\delta_3\beta_2\beta_0 - 
3\delta_1\beta_2^2 \\[0.1in]
\epsilon_3 & = & 4\left( \delta_4\beta_0^2 - \delta_0\beta_2^2\right) \\[0.1in]
\epsilon_2 & = & 3\delta_3\beta_0^2 - 2\delta_1\beta_2\beta_0 \\[0.1in]
\epsilon_1 & = & 2\delta_2\beta_0^2 - 4\delta_0\beta_2\beta_0 \\[0.1in]
\epsilon_0 & = & \delta_1\beta_0^2
\end{array} \qquad .
$$

Noticing that the quantity $\beta_2u_0^2 + \beta_0$ is positive for
all $u_0$, we can now state a numerical algorithm to solve for the
minimizing control $u_0^*$. First, determine the possible extrema via
the first-derivative condition of
\begin{equation}
\delta_5\omega^5 + \delta_4\omega^4 + \delta_3\omega^3 + \delta_2\omega^2 
+ \delta_1\omega + \delta_0  = 0 \qquad ,
\label{eq:LQGRex2poly1}
\end{equation}
which we know will be satisfied by at most five distinct, and in
general complex, values $\omega_1,\ldots,\omega_5$. We can immediately
disregard the $\omega_i$s that have nonzero imaginary components; if
multiple real-valued $\omega_i$s persist, we then apply the
second-derivative condition for each such $\omega_i$ and only retain the
ones for which the second-derivative condition is satisfied i.e., 
\begin{equation}
\epsilon_8\omega^8 + \epsilon_6\omega^6 + \epsilon_5\omega^5 +
\epsilon_4\omega^4 + \epsilon_3\omega^3 + \epsilon_2\omega^2 + 
\epsilon_1\omega + \epsilon_0 > 0 \qquad .
\label{eq:LQGRex2poly2}
\end{equation}
If no valid $\omega_i$ remains, then we conclude there is no
minimizing control $u_0^*$; if just one valid $\omega_i$ remains, then
we conclude $u_0^* = \omega_i$; if multiple $\omega_i$s remain, then
we choose the one for which $Q_0\left(P_{x_0|I_0},\omega_i\right)$ is
the smallest. If there is a ``tie,'' then we either choose the
$\omega_i$ that is furthest from the value of $m_{0|0}$ or, if this is
also a ``tie,'' then simply the smallest $\omega_i$. The fist plot in
Fig.~\ref{fig:lqgrEx1} is an example of when such a double-tie can arise,
where there is perfect symmetry of $Q_0$ about the $u_0=0$ axis. Here,
our algorithm would conclude with $u_0^* < 0$.

As a sanity check, consider again the case when $\gamma$ approaches
zero. In (\ref{eq:LQGRex2poly1}), numerator coefficients
$\delta_5,\ldots,\delta_2$ all approach zero and $\delta_1$ approaches
$2(r+b^2\kappa_1)\beta_0$, resulting in the single root
$$
\omega_1 = -\frac{\delta_0}{\delta_1} = -\frac{2ab\kappa_1m_{0|0}\beta_0}
{2(r+b^2\kappa_1)\beta_0} = -\frac{ab\kappa_1}{r+b^2\kappa_1} m_{0|0} \qquad .
$$
It is real-valued, so even without proceeding to the second
derivative test, the above algorithm concludes this is the minimizing
control. It, of course, again agrees with the standard LQGR solution.
Continuing nonetheless, in (\ref{eq:LQGRex2poly2}), numerator
coefficients $\epsilon_8,\epsilon_6,\ldots,\epsilon_1$ all approach
zero and $\epsilon_0$ approaches $2(r+b^2\kappa_1)\beta^3$, which is
indeed positive because we know $r>0$, $\kappa_1 \geq 0$ and $\beta >
0$.

\begin{figure}[t!]
\begin{center}
  {\bf Parameters:} $(m_x,\sigma_x,a,b,\sigma_w,c,d,\sigma_v,\gamma) =
  (0,1,1,1,1,1,0,1,10)$, \quad $(t,r) = (1,1)$ \\[0.2in]
\resizebox{0.75\textwidth}{!}{\includegraphics{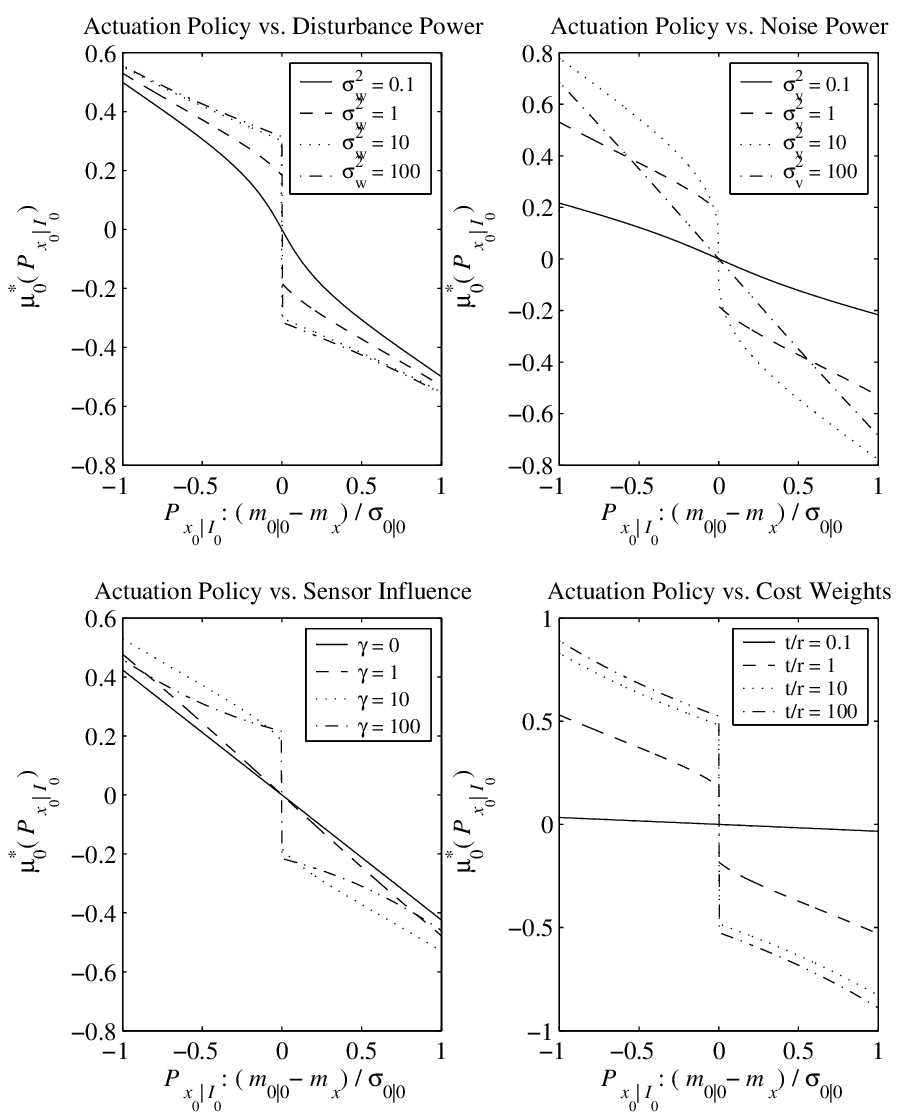}}
\caption{Sensitivity of Optimal Actuation Policy
With Direct Measurement Control}
\label{fig:lqgrEx2}
\end{center}
\end{figure}

We close the discussion of this example with a sensitivity analysis of
the optimal actuation strategy $\mu^*_0$, mapping a Gaussian
probabilistic state estimate $P_{x_0|I_0}$ parameterized by mean
$m_{0|0}$ and variance $\sigma^2_{0|0}$ to the minimizing control
$u^*_0$. Figure~\ref{fig:lqgrEx2} shows the output of this analysis,
where each individual plot varies one model parameter while holding all
others fixed at their nominal values. Thus, each individual curve
corresponds to a specific instance of all model parameters and is
generated by sweeping the conditional mean $m_{0|0}$ over $\pm 1$
standard deviation $\sigma_{0|0}$ from its predicted value of $m_x$, in
effect sweeping over a most-likely subset of possible measurements
$z_0$. For each sample $m_{0|0}$, the minimizing control $u_0^*$ is
computed by the above algorithm based on (\ref{eq:LQGRex2poly1}) and
(\ref{eq:LQGRex2poly2}).  We vary the (i) system disturbance power
captured by parameter $\sigma^2_w$, (ii) measurement noise power
captured by parameter $\sigma^2_v$, (iii) degree of sensor influence
captured by parameter $\gamma$, and (iv) relative weightings in the
quadratic cost function captured by the ratio of parameters $t$ and
$r$. In each plot, note that one of the values depicted for the variable
parameter corresponds to its nominal value (e.g., in the upper-left
plot, the dashed line corresponds to $\sigma_w^2 = 1$).

Common to all of the plots in Fig.~\ref{fig:lqgrEx2} is a potential
discontinuity in the minimizing control $u_0^*$ around the point that
the conditional mean $m_{0|0}$ is near its predicted value $m_x$. This
discontinuity is more apparent for (i) larger values of disturbance
power, (ii) larger values of noise power, (iii) larger values of sensor
influence, and (iv) larger values of cost weights. Note that we have
chosen our initial state estimate to be zero-mean, so when the estimate
$m_{0|0}$ is near its predicted value, then the expected quadratic
penalty in the state is near zero. Moreover, the disturbance variable
$w_0$ is also zero-mean by assumption. In the standard LQGR formulation,
the solution of which corresponds to the $\gamma=0$ curve in the
lower-left plot, these properties altogether imply that the minimizing
control in this case is also zero---that is, the state is already
regulated and therefore a non-regulated control can be of no
benefit. However, with an opportunity for sensor management, the
discontinuity represents the notion that applying a non-regulated
control at stage $0$ is desirable even if the state appears to be
regulated at stage $0$. The combination of potential error in the state
estimate at stage $0$ and the stochastic disturbance in the system
equation allows the controller to anticipate that the state is unlikely
to remain regulated for even a single stage. Therefore, the
discontinuity represents the controller trading off the undesirability
of purposefully exciting the state at stage $0$ versus the desirability
of a more accurate assessment of the true state in the next
stage. Indeed, we expect this tradeoff to be most acute when there are
large levels of uncertainty from stage-to-stage (i.e., larger values of
disturbance/noise powers), potential for large reductions in measurement
uncertainty per unit control (i.e., larger values of $\gamma$) or state
regulation is far more important than control regulation (i.e., larger
ratios $t/r$). This qualitative assessment, matching intuition, is
precisely quantified by the LQGR formulation with direct measurement
control. 

As a final aside, note that the optimal cost-to-go function is now
numerically computable according to
$$
J_0(P_{x_0|I_0}) = Q_0(P_{x_0|I_0},u_0^*) \qquad .
$$ Its expected value over the conditional Gaussian distribution with
mean $m_{0|0}$, itself a function of random variable $z_0$, and variance
$\sigma^*_{0|0}$ results in some function of the measurement
$z_0$. Iterating expectations, or then also taking the expected value
with respect to the Gaussian distribution of $z_0$, results in the
optimal expected total cost $J(\phi_2,\pi_2)$. It is conceivable to
perform these computations via a combination of analysis and numerical
methods, similar to the manner in which the calculations were carried
out for the optimal actuation policy $\mu_0^*$. This then allows one to
generate sensitivity curves for the optimal expected cost $J$, analogous
to those presented in Fig.~\ref{fig:mrEx4} for the machine repair
example, and draw similar conclusions about the kinds of LQGR
circumstances for which a sensor management scheme is indeed ``worth its
cost.''

\section{Conclusions and Extensions \label{Conclu}}
We have framed the opportunity for sensor management, motivated by many
modern engineering applications, as a stochastic control problem with
imperfect state information. We summarized the computational machinery
available for such problems and, under the assumption that optimality is
still our goal, reviewed the most fundamental concepts that allow the
transformation of these generally applicable problem formulations to
implementable solutions. It is worth emphasizing that an entire body of
literature is devoted to making this computational machinery even more
applicable in practice \cite{BeT96:NeuDP} via systematic approximation, but
these methods were not considered here. Rather, our modest focus was to
clearly illustrate how stochastic optimal control formulations with
imperfect state information do in fact encompass sensor management
problems. Had we introduced approximation from the start, all of our
``surprising'' conclusions would necessarily be clouded by suspicion of
approximation fidelity.

Restricted to optimality, we then considered two distinct classes of
well-studied stochastic control problems that turn out to admit exact
solution: a PO-MDP formulation and a LQGR formulation. In the former,
the standard model readily permitted the opportunity for sensor
management, and we studied in detail an extension of the classic
machine-repair scenario. In the LQGR formulation, however, the standard
formulation had to be extended to capture the opportunity for sensor
management. The first proposed extension, motivated by the PO-MDP
formulation, included a notion of measurement scheduling from a finite
set of sensors and we showed that augmenting off-line computation with
the solution to a special (deterministic) dynamic program reduced the
problem to the standard LQGR formulation. The second extension, however,
allowed direct control of future measurement noise, in a sense allowing
measurement selection from an infinite set of sensors. The computational
ramifications of such an approach were far more severe, yielding a
nonlinear stochastic control problem for which the many computational
simplifications inherent to LQGR problems no longer apply. These points
were concretely illustrated in a simplest instance of the extended LQGR
model, for which it was shown that the solution agreed with some
qualitative conclusions one might conjecture based on intuition.

There are several ways in which the work in this paper could be
extended. Based on the LQGR discussion, the applicability of existing
nonlinear control methods can be investigated.  In problems for which an
optimal solution is hopeless, approximate dynamic programming methods
can be explored. Most emerging concepts of sensor management involve
decision problems that are supported by so-called sensor networks,
introducing notions of local processing, costly communications,
distributed algorithms and other dimensions of the general class of
decentralized control problems. Of course, all of these extensions make
an already difficult computational problem even more difficult \cite{TsA85:CoDec}; thus,
the quest for improved sensor management solutions promises to challenge
all researchers, mathematicians, scientists and engineers alike, for
many years to come.

\bibliography{../bibtex/Textbooks,../bibtex/SensorManagement}

%\appendix

%\section{MATLAB Code}
%
%(Incomplete): Computational methods for PO-MDP example and
%Case 2 of LQGR example. Written already, but needs to be modularized
%better for inclusion here.

\end{document}